\documentstyle[aps,preprint,floats,psfig]{revtex}

\tighten

\begin{document}

\preprint{\scriptsize{To appear in Phys.\ Rev.\ C {\bf 56} (Oct. 1997) 
   --- {\bf nuclth/9704057} [1808:1200]}}
\draft

\title{\vspace{1cm} 
Gauge-invariant theory of pion photoproduction\protect \\
with dressed hadrons}
\author{Helmut Haberzettl}
\address{Center for Nuclear Studies, Department of Physics,\protect \\
The George Washington University, Washington, D.C. 20052\\
and\\
Institut f\"ur Kernphysik (Theorie), Forschungszentrum J\"ulich GmbH,%
\protect \\
D-52425 J\"ulich, Germany}
\date{21 April 1997}
\maketitle

\begin{abstract}
Based on an effective field theory of hadrons in which quantum
chromodynamics is assumed to provide the necessary bare cutoff functions, a
gauge-invariant theory of pion photoproduction with fully dressed nucleons
is developed. The formalism provides consistent dynamical descriptions of 
$\pi N\rightarrow \pi N$ scattering and $\gamma N\rightarrow \pi N$
production mechanisms in terms of nonlinear integral equations for fully
dressed hadrons. Defining electromagnetic currents via the gauging of
hadronic $n$-point Green's functions, dynamically detailed currents for
dressed nucleons are introduced. The dressed hadron currents and the pion
photoproduction current are explicitly shown to satisfy gauge invariance in
a self-consistent manner. Approximations are discussed that make the
nonlinear formalism manageable in practice and yet preserve gauge invariance.
This is achieved by recasting the gauge conditions for all contributing
interaction currents as continuity equations with ``surface'' terms for the
individual particle legs coming into or going out of the hadronic
interaction region. General procedures are given that approximate any type
of (global) interaction current in a gauge-invariance-preserving manner as a
sum of single-particle ``surface'' currents. It is argued that these
prescriptions carry over to other reactions, irrespective of the number or
type of contributing hadrons or hadronic systems.
\end{abstract}


%%%%%%%%%%%%%%%%%%%%%%%%%%%%%%%%%%%%%%%%%%%%%%%%%%%%%%%%%%%%%%%%%%%%%%%%

\narrowtext

%%%%%%%%%%%%%%%%%%%%%%%%%%%%%%%%%%%%%%%%%%%%%%%%%%%%%%%%%%%%%%%%%%%%%%%%

\section{Introduction}

While it is generally accepted that quantum chromodynamics (QCD) provides a
basic framework for all reactions involving strongly interacting particles,
we still seem to be very far away from implementing QCD in practical
calculations aimed at describing the findings of experiments from low to
intermediate energies of up to a few GeV. Instead, effective
fieldtheoretical descriptions in terms of purely hadronic degrees of freedom
are usually employed, where QCD is assumed to provide the justification for
the parameters or cutoff functions assumed in the various approaches.
Central to our ability to describe experiments in these effective terms ---
and, most importantly, central to our ability to tell when this effective
approach would no longer be applicable --- is a detailed understanding of
the most basic hadronic interactions, namely the reaction dynamics of
nucleons, pions, and photons.

It is the purpose of the present paper to provide a comprehensive
theoretical description of the production of pions due to the interactions
of incident photons with nucleons. The history of such descriptions goes
back to the fifties, and indeed many of the (global) basic relations
have been well-known for about forty years (see \cite{WTI} and, in particular, 
\cite{kazes}, and references therein). In recent years, the attention has
focused on approaches attempting to take into account the fact that all
hadrons involved in the reaction have an internal structure \cite
{gross,afnan1,ohta,friar,coester,afnan,banerjee}.

The present investigation would like to add to the latter approaches by
providing a detailed theoretical framework for the gauge-invariant interactions
of physical
--- i.e., fully dressed --- hadrons with photons. The description is given
in terms of an effective field theory where the (at present, in detail
unknown) quark and gluon degrees of freedom are parametrized by the bare
quantities of the effective Lagrangian. Since many of the most basic
relations governing the interactions of hadrons and photons relate
electromagnetic to purely hadronic entities (the most important example
being the Ward-Takahashi identities \cite{WTI}), it seems obvious that a
comprehensive formulation can only be achieved if the purely hadronic sector
of the problem is treated completely consistently with the subsequent
electromagnetic interaction. Therefore, in order to be able to develop the
present description from the ground up, we will, in Sec.\ II, provide a
recapitulation of the interactions of nucleons and pions that will form the
basis for the ensuing investigations regarding the photoproduction of pions.
The formulation given here for $\pi N\rightarrow \pi N$ is a nonlinear one
where the full solution of the hadronic scattering problem couples back into
the driving term of the reaction. Despite the practical complications
entailed by nonlinear $\pi N$ scattering equations, we feel very strongly
that, from both formal and practical points of view, the nonlinear approach
is better suited to exhibit the true dynamics of the interaction, and as a
consequence, if a problem needs to be treated as a linear one because of
practical considerations, starting from the full nonlinear set of equations
usually will suggest an approximation strategy closer to the true underlying
physical mechanisms than starting from linearized assumptions from the very
beginning.

The physical currents for all processes contributing to the pion production 
amplitude are derived here via their corresponding hadronic $n$-point Green's 
functions. However, instead of the usual procedure (see, e.g., 
\cite{ohta,afnan}) of employing minimal substitution and a subsequent 
functional derivative with respect to the electromagnetic field $A^{\mu}$, 
we introduce an equivalent, but practically simpler, mathematical operation 
called a ``gauge derivative'' which allows one to obtain currents directly 
from the momentum-space versions of the respective Green's 
functions.\footnote{%
After completion of the present work, the author became aware of recent e-prints
by Kvinikhidze and Blankleider \cite{kbweb} which suggest that they employ a
similar method when gauging hadronic spectator equations.
However, at present, no technical details of their formalism are 
available. Therefore, the exact nature of the relation to the 
gauge-derivative method introduced here
remains to be investigated.} The
technical details of this are explained in the Appendix. Using the hadronic 
results of Sec.\ II as a starting point for the definitions of all relevant 
$n$-point Green's functions, we arrive in Sec.\ III at a consistent formulation 
of the pion photoproduction process where all reaction mechanisms are given in 
terms of fully dressed hadronic propagators and vertex functions. The 
electromagnetic current for the physical (i.e., dressed) nucleon, in 
particular,  contains all contributions from the nucleon's self-energy. 
Again, the results are found to be highly nonlinear and, hence, the proof 
of gauge invariance of the formalism can only be one of self-consistency, 
where we show, in Sec.\ IV, that assuming the validity of the Ward-Takahashi 
identities \cite{WTI} will provide a gauge-invariant pion photoproduction 
amplitude which in turn will allow us to get back the Ward-Takahashi identities 
for the dressed hadron currents in a self-consistent manner. In performing this 
proof, we observe that the 
gauge-invariance conditions for all current mechanisms inside of a hadronic 
interaction region --- as opposed to pieces arising from the photon's interaction
 with external legs of the reaction --- take the form of continuity equations with
 ``surface'' terms. Using the example of the electromagnetic current associated 
with the bare vertex as a starting point, in Sec.\ V, it is shown in the concluding 
Sec.\ VI how one can turn this into a consistent approximation scheme that allows 
one to preserve the gauge invariance of the formalism even if one chooses to 
linearize the problem by altering appropriately the driving terms for the 
underlying $\pi N\rightarrow\pi N$ scattering problem. This approximation scheme 
is one of the most important practical results of the present investigation. While, 
at first glance, some of the resulting expressions might look similar to Ohta's 
\cite{ohta} way of dealing with extended nucleons, it will be seen that there are 
considerable differences in detail which even leads to a simpler implementation in 
practice (with, for example, none of the problems pointed out in 
Ref.\ \cite{banerjee}). Moreover, we also argue in 
Sec.\ VI that the approximation scheme suggested here for the $\pi N$ system can 
also be used for other reactions irrespective of the type or number of contributing 
hadrons.

%%%%%%%%%%%%%%%%%%%%%%%%%%%%%%%%%%%%%%%%%%%%%%%%%%%%%%%%%%%%%%%%%
\section{${\pi }N$ Dynamics}

As explained in the Introduction, the basis of a consistent treatment of the
reaction dynamics of pions, nucleons, and photons is a complete formulation
of the purely hadronic sector of the problem, i.e., $\pi N\rightarrow \pi N$%
. To simplify the presentation, we will in the following only consider
pion-nucleon interactions proceeding through the $P_{11}$ and the $P_{33}$
channels. Moreover, the only bound or resonance states taken into account in
these channels will be the nucleon, $N$, and the delta, $\Delta $,
respectively. Also, we will allow for $\pi \pi $ interactions that give rise
to $\sigma $ and $\rho $ degrees of freedom. Other meson contribution will
not be treated here. We thus keep the reaction dynamics simple enough to be
amenable to a concise presentation, yet sufficiently detailed to avoid being
trivial. However, we emphasize that none of these restrictions are
essential; a more elaborate dynamical picture can easily be obtained
following the lines presented here.

Many, if not most, aspects of the following treatment of hadronic $\pi N$
dynamics are well known 
(see Refs.\ \cite{pin1,pin2,pin3,pin4} and references therein). 
Nevertheless, we feel that it is necessary to
recapitulate them here in order to provide a completely consistent
background against which the investigations of Secs.\ III and IV regarding
the gauge invariance of the pion photoproduction amplitude needs to be seen.

%%%%%%%%%%%%%%%%%%%%%%%%%%%%%%%%%%%%%%%%%%%%%%%%%%%%%%%%%%%%%%%%%%%%%%%%

\subsection{The dressed nucleon propagator}

Generically, the bare Lagrangian for the nucleon is given by \cite{weinberg}

\begin{equation}
{\cal L}=\bar{\psi}_{B}(i{\partial \!\!\!/}-m_{B})\psi _{B}-V_{B}(\bar{\psi}%
_{B},\psi _{B})\,\,\,,  \label{eq1}
\end{equation}
where the index $B$ stands for ``bare'' and $V_{B}$ includes all the
interactions relevant for the nucleon; in general, $V_{B}$ will also contain
other fields. We assume here that the bare quantities have been obtained
from QCD, subsuming the quark and gluon degrees of freedom that cannot be
described by purely hadronic dynamics. At the hadronic level, for the
nucleon the only remnants of this procedure are the bare mass, $m_{B}$, and
a bare vertex function for $N\rightarrow N\pi $ , 
\begin{equation}
F_{B}(p^{\prime },p)=\,g_{B}\,G_{5}\,f(p^{\prime },p)\,\,\,,  \label{eq1-1}
\end{equation}
where $g_{B}$ is the bare coupling constant and $G_{5}$ the coupling
operator. The function $f(p^{\prime },p)$ --- which in practical terms is
free input, since the QCD problem has not (yet) been solved --- provides the
necessary cutoff to make all integrals convergent. For $G_{5}$ we are only
going to consider pseudoscalar, $G_{5}=\gamma _{5}$, or pseudovector, $%
G_{5}=\gamma _{5}({p\!\!\!/}-{p\!\!\!/^{\prime })}/2m$, couplings. (Again, a
more general coupling structure can easily be accommodated following the
outline given here.)

Note that the $\pi NN$ vertex conserves four-momentum and the independent
variables chosen in Eq.\ (\ref{eq1-1}) are the four-momenta of the incoming
and outgoing nucleons, $p$ and $p^{\prime }$, respectively. As we shall see
in Sec.\ V, this will have some bearing on the investigation of gauge
invariance of the pion photoamplitude.

Renormalizing the Lagrangian in the usual way \cite{weinberg} by

\begin{eqnarray}
m_{B}+\delta m &=&m\,\,\,,  \label{eq2} \\
\psi _{B}\frac{1}{\sqrt{Z}} &=&\psi \,\,\,,  \label{eq3} \\
V_{B}(\sqrt{Z}\bar{\psi}_{B},\sqrt{Z}\psi _{B}) &=&V(\bar{\psi},\psi )\,\,\,,
\label{eq4}
\end{eqnarray}
yields

\begin{equation}
{\cal L}=Z\,\bar{\psi}(i{\partial \!\!\!/}-m)\psi +Z\delta m\,\bar{\psi}\psi
-V(\bar{\psi},\psi )\,\,\,.  \label{eq5}
\end{equation}
The inverse of the corresponding momentum-space propagator is then given by 
\begin{equation}
S^{-1}(p)=({p\!\!\!/}-m)Z+Z\delta m-\Sigma ({p\!\!\!/})\,\,\,,  \label{eq6}
\end{equation}
where the self-energy ${\Sigma ({p\!\!\!/})}$ is the sum of all hadronic
one-particle irreducible loops that can be constructed with $V(\overline{%
\psi },\psi )$, i.e., ${\Sigma ({p\!\!\!/})}$ contains all graphs that
cannot be factorized by simply cutting one internal nucleon line. Choosing
the renormalization parameters as

\begin{eqnarray}
Z\delta m &=&\Sigma (m)\,\,\,,  \label{eq7} \\
Z &=&1+\left. \frac{d{\Sigma ({p\!\!\!/})}}{d{p\!\!\!/}}\right| _{{p\!\!\!/}%
=m}\,\,\,,  \label{eq8}
\end{eqnarray}
assures that the fully dressed propagator (\ref{eq6}) has a pole at the
nucleon mass, $m$, with a unit residue.

Introducing a bare propagator by 
\begin{equation}
S_{0}^{-1}(p)=({p\!\!\!/}-m_{B})Z=({p\!\!\!/}-m)Z+\Sigma (m)\,\,\,,
\label{eq9}
\end{equation}
i.e., 
\begin{equation}
S^{-1}(p)=S_{0}^{-1}(p)-\Sigma ({p\!\!\!/})\,\,\,,  \label{eq9-9}
\end{equation}
the bare and dressed propagators are easily seen to be related by

\begin{equation}
S(p)=S_{0}(p)+S_{0}(p)\Sigma ({p\!\!\!/})\,\,S(p)\,\,\,.  \label{eq10}
\end{equation}
The bare propagator is the one to be associated with the bare vertex of Eq.\ (%
\ref{eq1-1}); in the renormalized scheme the bare s-channel pole term
contributing to the $T$-matrix for $\pi N\rightarrow \pi N$ arising from the
Lagrangian (\ref{eq1}) is given as\footnote{%
Throughout this paper, the bra-ket notation is used as a quick way to see
whether one deals with $N\pi \rightarrow N$ transitions or $N\rightarrow
N\pi $.}

\begin{eqnarray}
V_{0} &=&\left| F_{B}\right\rangle \,\frac{1}{{p\!\!\!/}-m_{B}}%
\,\left\langle F_{B}\right|  \nonumber \\
&=&Z\left| F_{B}\right\rangle S_{0}\left\langle F_{B}\right|  \nonumber \\
&=&\left| F\right\rangle S_{0}\left\langle F\right| \,\,\,,  \label{eq11}
\end{eqnarray}
where the (renormalized) bare vertex is given by, cf. Eq.\ (\ref{eq1-1}), 
\begin{equation}
F(p^{\prime },p)\,\,=g_{0}\,G_{5}\,f(p^{\prime },p)\,\,\,,\,\,  \label{eq12}
\end{equation}
with $g_{0}=\sqrt{Z}g_{B}$, in accordance with the renormalized interaction (%
\ref{eq4}). (Note that in keeping with our simplifying assumptions we do not
consider more than one s-channel term per $\pi N$ channel.)

The formalism given so far assumes that there exists a pole in the baryon
channel at hand, which applies only to the dressed nucleon in the $P_{11}$
channel. For the $P_{33}$ channel, with the $\Delta $ resonance, the bare
vertex (\ref{eq1-1}) and the renormalization conditions (\ref{eq7}) and (\ref
{eq8}) need to be modified accordingly. The conditions for a resonance,
rather than a bound state, are obtained by replacing the self-energy by its
real part and the bound-state mass by the real part of the resonance
position. Also, the spin--1/2 propagator needs to be replaced by one for
spin--3/2. Since it has no direct bearing on the treatment given here, we
will not go into further details.

%%%%%%%%%%%%%%%%%%%%%%%%%%%%%%%%%%%%%%%%%%%%%%%%%%%%%%%%%%%%%%%%%%%%%%%%

\subsection{The pion}

In complete analogy to the preceding nucleon treatment, one may derive a
dressed pion propagator $\Delta (q)$ given by \cite{weinberg} 
\begin{equation}
\Delta ^{-1}(q)=(q^{2}-\mu ^{2})Z_{\pi }+\Pi (m_{\pi }^{2})\,-\Pi
(q^{2})\,\,,  \label{eq13}
\end{equation}
where $m_{\pi }$ is the (physical) pion mass, $\Pi (q^{2})$ the pion
self-energy, and $Z_{\pi }$ the renormalization of the pion field. However,
for the present report, the details of this dressing are not important; we
merely require that $\Pi (q^{2})$ be such that the dressed pion propagator
allows for a Ward-Takahashi identity \cite{WTI,weinberg} in analogy to what
will be derived for the nucleon. We emphasize that this can indeed be
achieved quite straightforwardly by applying to the pion the corresponding
steps outlined below for the nucleon. For all practical purposes, we may
even replace $\Delta $ by the bare pion propagator $\Delta _{0}$, 
\begin{equation}
\Delta ^{-1}(q)\rightarrow \Delta _{0}^{-1}(q)=q^{2}-m_{\pi }^{2}\,\,\,,
\label{eq13a}
\end{equation}
without changing any of the findings reported here.

%%%%%%%%%%%%%%%%%%%%%%%%%%%%%%%%%%%%%%%%%%%%%%%%%%%%%%%%%%%%%%%%%%%%%%%%

\subsection{$\pi N$ scattering}

In a graphical picture, for each of the two channels considered here for $%
\pi N$ scattering, there will be infinitely many graphs contributing to each 
$T$-matrix arising from the nucleon Lagrangian (\ref{eq1}) and its delta
counterpart. Summing up subclasses of graphs at the $\pi N$-reducible level
produces graphs that can be classified as to how many separate graph
fragments one obtains by cutting across a pair of {\it fully dressed} $\pi $
and $N$ propagators. Denoting the pair of $\pi N$ propagators formally by $%
G_{0}$, 
\begin{equation}
G_{0}(P)=S(p)\circ \Delta (q)\,\,\,,  \label{eq14}
\end{equation}
where $\circ $ denotes the convolution\footnote{%
By convolution, we mean that dynamically the pion and nucleon propagators $%
\Delta (q)$ and $S(p)$ are independent, except for the fact that they share
a conserved total four-momentum $P=p+q$. In loops every convolution is
associated with an integration $i\int d^{4}q$.} of the propagators with
fixed total four-momentum $P=p+q$, generically each of these summed-up
graphs can thus be written as

\begin{equation}
{\rm graph}=AG_{0}BG_{0}C...G_{0}D\,\,\,.  \label{eq15}
\end{equation}
where none of the $\pi N$-irreducible pieces $A,B,C,D$, etc. can be written
in a similar way. Denoting the sum of all of these pieces by $U$, adding to
it the one-particle reducible bare s-channel term $V_{0}$ of Eq.\ (\ref{eq11}%
) and denoting the result by $V$, i.e.,

\begin{eqnarray}
V &=&V_{0}+U  \nonumber \\
&=&\left| F\right\rangle S_{0}\left\langle F\right| +U\,\,\,,  \label{eq17}
\end{eqnarray}
generically both reactions then are formally given by

\begin{eqnarray}
T &=&V+VG_{0}V+VG_{0}VG_{0}V+...  \nonumber \\
&=&V+VG_{0}T\,\,\,.  \label{eq16}
\end{eqnarray}
The Bethe--Salpeter \cite{BSE} integral equations thus derived for the
nucleon and delta channels are coupled via their respective driving terms $V$%
. Denoting the $T$-matrices by $T_{N}$ and $T_{\Delta }$, with driving terms 
$V_{N}$ and $V_{\Delta }$, one has

\begin{eqnarray}
T_{N} &=&V_{N}[T_{\Delta }]+V_{N}[T_{\Delta }]\,G_{0}T_{N}\,\,\,,  \nonumber
\\
T_{\Delta } &=&V_{\Delta }[T_{N}]+V_{\Delta }[T_{N}]\,G_{0}T_{\Delta }\,\,\,,
\label{eq16-1}
\end{eqnarray}
where $V[T]$ denotes the functional dependence.

The effect of the bare, one-particle reducible pole-term $V_{0}$ on the $T$%
-matrix is easily seen by defining an auxiliary entity by a Bethe--Salpeter
equation with $V_{0}$ removed, 
\begin{equation}
X=U+UG_{0}X\,\,\,,  \label{eq18}
\end{equation}
and applying the well-known two-potential formula \cite{goldberger} to Eq.\ (%
\ref{eq16}) using (\ref{eq17}) and (\ref{eq18}). This produces \widetext
\begin{equation}
T=(1+XG_{0})\left| F\right\rangle \,\frac{1}{S_{0}^{-1}-\left\langle
F\right| (G_{0}+G_{0}XG_{0})\left| F\right\rangle \,}\left\langle F\right|
(G_{0}X+1)+X  \label{eq19}
\end{equation}
\noindent or

\begin{equation}
T=\left| \Gamma \right\rangle S\left\langle \Gamma \right| +X\,\,\,;
\label{eq20}
\end{equation}
in other words, the $T$-matrix has been split into its pole contribution and
a nonpolar piece described by $X$. The self-energy of the dressed
propagator $S,$ formally given by Eq.\ (\ref{eq9-9}), has been obtained here
as 
\begin{equation}
\Sigma = \left\langle F\right| (G_{0}+G_{0}XG_{0})\left| F\right\rangle 
  = \left\langle F\right| G_{0}\left| \Gamma \right\rangle \,=\left\langle
\Gamma \right| G_{\,0}\left| F\right\rangle \,\,\,,  \label{eq21}
\end{equation}
where 
\begin{equation}
\left| \Gamma \right\rangle = \left( 1+XG_{0}\right) \left| F\right\rangle 
          =\left| F\right\rangle +UG_{0}\left| \Gamma \right\rangle \,\,\,
\label{eq22}
\end{equation}
\narrowtext
\noindent defines the dressed vertex function. In view of the unit residue of $S$, we
may assume here without loss of generality that $\Gamma $ is already
properly normalized, i.e., that the coupling constant associated with the
dressed vertex is the physical $\pi NN$ vertex constant; in practical terms
this is achieved by choosing the bare coupling $g_{0}$ accordingly.

Obviously only the present procedure lends substance to the formal dressing
discussed in Sec.\ II.A\@. The summary of the relevant equations in Fig.\ 1 
\begin{figure}
\centerline{\psfig{file=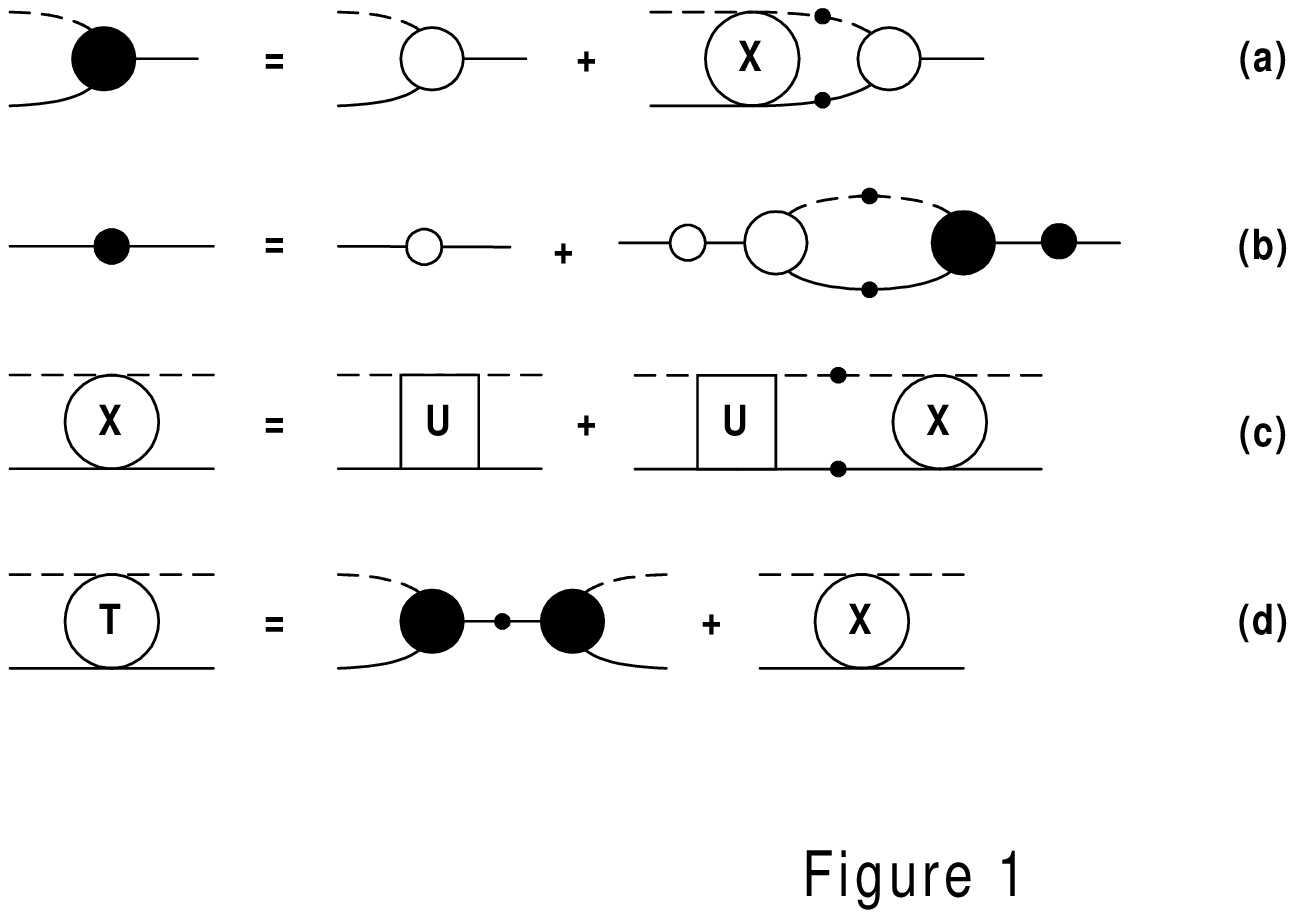,width=5in,clip=}}
\vspace{3mm}
\caption{Summary of coupled, nonlinear equations for {$\pi N\rightarrow \pi
N $}: (a) Dressed vertex, Eq.\ (\ref{eq22}); (b) dressed propagator, 
Eq.\ (\ref{eq10}); (c)
nonpolar amplitude $X$, Eq.\ (\ref{eq18}); (d) full $T$-matrix, 
Eq.\ (\ref{eq20}). Open
circles (without descriptive letters) denote bare quantities, whereas solid 
circles depict the respective fully dressed ones.}
\end{figure}
also
shows that the formulation in terms of dressed hadrons leads to a {\it %
nonlinear} system of equations where --- ignoring pion dressing for
simplicity --- Eqs.\ (\ref{eq10}), (\ref{eq18}), and (\ref{eq22}) feed into
each other in a very complex manner. Solving this set of nonlinear equations
presents a formidable task in practical terms. Perhaps it is for this reason
that nonlinear formulations of the pion-nucleon problem are often avoided.
However, the $\pi N$ scattering problem {\it is} nonlinear. We therefore
would like to advocate that substantial insights can be gained from first
formulating the problem in its full complexity and then implementing
approximations that make the problem manageable, instead of linearizing the
problem from the very beginning. The relative simplicity of the
gauge-invariance investigation of Sec.\ IV might be taken as a case in point
for this view.

%%%%%%%%%%%%%%%%%%%%%%%%%%%%%%%%%%%%%%%%%%%%%%%%%%%%%%%%%%%%%%%%%%%%%%%%

\subsection{The nonpolar driving term}

By construction, the nonpolar driving term $U$ is two-particle irreducible;
in other words, every internal cut across an entire diagram must necessarily
cut across at least three particle lines. Moreover, as seen from Eqs.\ (\ref
{eq10}) and (\ref{eq22}), the complete dressings of propagators and vertices
happen at the one- and two-particle {\it reducible} levels, respectively.
Therefore, since $U$ is at most three-particle reducible, without loss of
generality all contributions to $U$ can be taken as being constructed in
terms of fully dressed propagators and vertices.

Some of the lowest-order contributions thus are readily seen to be given by 
\begin{eqnarray}
U_{N} &=&\left\langle \Gamma _{N}\right| \overline{S}_{N}\left| \Gamma
_{N}\right\rangle +B_{N}+...  \nonumber \\
U_{\Delta } &=&\left\langle \Gamma _{\Delta }\right| \overline{S}_{\Delta
}\left| \Gamma _{\Delta }\right\rangle +B_{\Delta }+...  \label{eq24}
\end{eqnarray}
with both channels having similar box-type contributions $B_{N}$ and $%
B_{\Delta }$, i.e., 
\begin{eqnarray}
B &=&\left\langle \Gamma _{N}\right| \,\left( \overline{S}_{N}X_{N}\overline{%
S}_{N}\right) \circ \Delta \,\left| \Gamma _{N}\right\rangle  \nonumber \\
&&+\left\langle \Gamma _{N}\right| \,\left( \overline{\Delta }T_{\pi \pi }%
\overline{\Delta }\right) \circ S_{N}\,\left| \Gamma _{N}\right\rangle 
\nonumber \\
&&+\left\langle \Gamma _{\Delta }\right| \,\left( \overline{\Delta }T_{\pi
\pi }\overline{\Delta }\right) \circ S_{\Delta }\,\left| \Gamma _{\Delta
}\right\rangle  \nonumber \\
&&+\left\langle \Gamma _{N}\right| \,\left( \overline{S}_{N}T_{\Delta }%
\overline{S}_{N}\right) \circ \Delta \,\left| \Gamma _{N}\right\rangle
\,\,\,,  \label{eq25}
\end{eqnarray}
with the differences between $B_{N}$ and $B_{\Delta }$ being due to their total
spin and isospin values. $\Gamma _{N}$ and $\Gamma _{\Delta }$ here are the
dressed vertices for $\pi NN$ and $\pi N\Delta $, respectively. Figure 2
\begin{figure}
\centerline{\psfig{file=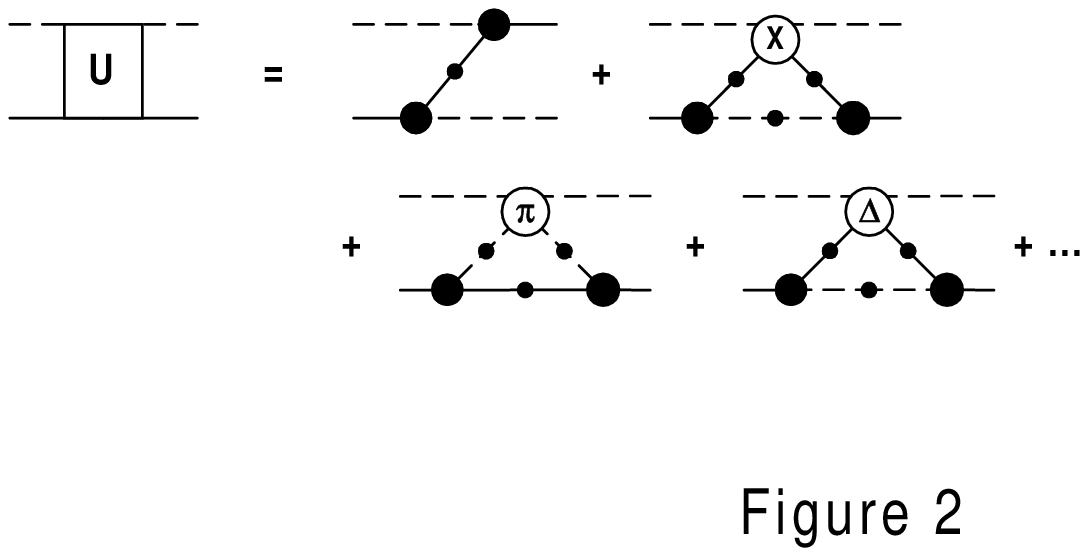,width=5in,clip=}}
\vspace{3mm}
\caption{Some of the lowest-order contributions for the driving term $U$ of
the nonpolar amplitude $X$ of Fig.\ 1(c). First graph on right-hand side is
the u-channel crossed counterpart of the s-channel graph of Fig.\ 1(d). The
other three box graphs depict intermediate scattering processes of the
nonpolar {$P_{11}$ $\pi N$} amplitude $X$ dressed by a pion, of the {$\pi
\pi $} amplitude dressed by a baryon (which subsumes both nucleon and
delta), and of the full {$P_{33}$ $\pi N$}
amplitude with pion dressing, respectively. All vertices and propagators
here are fully dressed.}
\end{figure}
depicts this structure. The respective first terms $\left\langle \Gamma
\right| \overline{S}\left| \Gamma \right\rangle $ here are the u-channel
counterparts of the pole terms $\left| \Gamma \right\rangle S\left\langle
\Gamma \right| $ of Eq.\ (\ref{eq20}). The bar over $S$ signifies that $S$ in
the context of $\left\langle \Gamma \right| \overline{S}\left| \Gamma
\right\rangle $ describes a baryon exchange across vertices and not a
loop-type matrix element (cf. first graph on the right-hand side of Fig.\ 2).
Furthermore, $U$ contains several box graphs where intermediate two-particle
scattering is dressed by a third particle: the two terms $\left\langle
\Gamma \right| \,\left( \overline{\Delta }T_{\pi \pi }\overline{\Delta }%
\right) \circ S\,\left| \Gamma \right\rangle $ describe $\pi \pi $
scattering dressed by a nucleon and a delta; again, the pion exchange (see
Fig.\ 2) is indicated by the bar over the pion propagator $\Delta $.
Intermediate $\pi N$ scattering proceeds through both $P_{11}$, $%
\left\langle \Gamma _{N}\right| \,\left( \overline{S}_{N}X_{N}\overline{S}%
_{N}\right) \circ \Delta \,\left| \Gamma _{N}\right\rangle \,$, and $P_{33}$%
, $\left\langle \Gamma _{N}\right| \,\left( \overline{S}_{N}T_{\Delta }%
\overline{S}_{N}\right) \circ \Delta \,\left| \Gamma _{N}\right\rangle $.
For the $P_{11}$ nucleon channel, however, the scattering involves only the
nonpolar contribution $X_{N}$ since the pole term provides a $\pi N$%
-reducible intermediate state, i.e., this contribution is generated through
an iteration of the u-channel crossed diagram. In keeping with the
simplifying assumptions made here, we have omitted intermediate scattering
processes like $\pi N\rightarrow \pi \Delta ,$ $\pi \Delta \rightarrow \pi
\Delta $, etc., and dressings by other mesons. However, there should be no
problem in writing down the corresponding contributions following the
procedure outlined here.

The driving term $U$ of the nonpolar amplitude $X$ is seen here to depend
not just on the $T$-matrix of the respective other channel, the way it was
written in Eq.\ (\ref{eq16-1}), but also on the complete solution of its own
channel, i.e., generically one can write Eq.\ (\ref{eq18}) as 
\begin{equation}
X=U[X]+U[X]\,G_{0}\,X\,\,\,.  \label{eq26}
\end{equation}
Without going into any details here, we would like to submit that this
nonlinear structure is of immediate practical consequence in that some of
the resonances usually considered as being independent actually originate
from dressed pole contributions of the intermediate $T_{\pi \pi }$ and/or $%
T_{\Delta }$ amplitudes.

%%%%%%%%%%%%%%%%%%%%%%%%%%%%%%%%%%%%%%%%%%%%%%%%%%%%%%%%%%%%%%%%%%%%%%%%

\section{The pion photoproduction amplitude}

In this Section, we will introduce a current for the dressed nucleon and the
corresponding pion-production current. In doing so, we will define an
exchange current, describing electromagnetic interactions during hadronic
exchange processes. The latter will be contributing to an all-encompassing
interaction current for the photon's action within the hadronic interaction
region.

%%%%%%%%%%%%%%%%%%%%%%%%%%%%%%%%%%%%%%%%%%%%%%%%%%%%%%%%%%%%%%%%%%%%%%%%

\subsection{The current of the dressed nucleon}

In a simplified picture, ignoring angular-momentum effects, the primary
dynamical change brought about by a photon entering a hadronic system is
that it deposits its four-momentum $k$ into a charged constituent of the
system thus changing the latter's momentum from $p\rightarrow p+k$. At its
most elementary level this is the {\it dynamical} basis for the usual
minimal substitution rule which in practical terms is often paraphrased as
``attach a photon to every momentum-dependent piece of a hadronic graph.''
In the Appendix, we define an operation, called the ``gauge derivative,''
that is similar in its effect to a functional derivative $\delta /\delta
A^{\mu }$, which allows one to formalize this procedure in a very simple
manner and derive current operators from hadronic $n$-point Green's
functions.

According to Eq.\ (\ref{Ncurr}) of the Appendix, the current of a nucleon is
given by

\begin{equation}
J^{\mu }(p^{\prime },p)=\{S^{-1}(p)\}^{\mu }\,\,\,,  \label{eq27}
\end{equation}
where $\{...\}^{\mu }$ is the gauge derivative defined in the Appendix, and $%
k=p^{\prime }-p$ is the momentum of the incoming photon. Using the rules of
the Appendix, we then find (omitting all momentum dependence for notational
clarity) \widetext
\begin{eqnarray}
J^{\mu } &=&\left\{ S_{0}^{-1}-\left\langle F\right| G_{0}\left| \Gamma
\right\rangle \right\} ^{\mu }  \nonumber \\
&=&Q_{N}Z\gamma ^{\mu }+\left\langle F^{\mu }\right| G_{0}\left| \Gamma
\right\rangle +\left\langle F\right| G_{0}g^{\mu }G_{0}\left| \Gamma
\right\rangle +\left\langle F\right| G_{0}\left| \Gamma ^{\mu }\right\rangle
\label{eq28}
\end{eqnarray}
where $F^{\mu }=-\{F\}^{\mu }$, $G_{0}g^{\mu }G_{0}=-\{G_{0}\}^{\mu }$, and $%
\Gamma ^{\mu }=-\{\Gamma \}^{\mu }$ result from attaching a photon to the
bare vertex, the two constituents of the $\pi N$ pair-propagator, and the
dressed vertex, respectively; see the Appendix for the corresponding
definitions and other technical details. Using Eq.\ (\ref{eq22}), one finds
for the dressed vertex, 
\begin{eqnarray}
\left| \Gamma ^{\mu }\right\rangle &=&\left| F^{\mu }\right\rangle +U^{\mu
}G_{0}\left| \Gamma \right\rangle +UG_{0}g^{\mu }G_{0}\left| \Gamma
\right\rangle +UG_{0}\left| \Gamma ^{\mu }\right\rangle  \nonumber \\
&=&(1+XG_{0})\left[ \,\left| F^{\mu }\right\rangle +g^{\mu }G_{0}\left|
\Gamma \right\rangle +U^{\mu }G_{0}\left| \Gamma \right\rangle \,\right]
-g^{\mu }G_{0}\left| \Gamma \right\rangle  \nonumber \\
&=&(1+XG_{0})\left| b^{\mu }\right\rangle -g^{\mu }G_{0}\left| \Gamma
\right\rangle  \nonumber \\
&=&\left| m^{\mu }\right\rangle -g^{\mu }G_{0}\left| \Gamma \right\rangle
\label{eq29}
\end{eqnarray}
where the exchange current $U^{\mu }=-\{U\}^{\mu }$ arises from the photon's 
interactions within the driving term $U$
(see subsection III.B for details) and

\begin{equation}
\left| b^{\mu }\right\rangle =\left| F^{\mu }\right\rangle +g^{\mu
}G_{0}\left| \Gamma \right\rangle +U^{\mu }G_{0}\left| \Gamma \right\rangle
\,\,\,,  \label{eq30}
\end{equation}
\begin{equation}
\left| m^{\mu }\right\rangle =\left( 1+XG_{0}\right) \left| b^{\mu
}\right\rangle = \left| b^{\mu }\right\rangle +UG_{0}\left| m^{\mu
}\right\rangle  \label{eq30-1}
\end{equation}
were used as abbreviations.  Note that the quantity 
$m^{\mu }$ appearing here has the same integral-equation kernel as the
nonpolar hadronic amplitude $X$ of Eq.\ (\ref{eq18}). Its relation to the
full photoproduction amplitude $M^{\mu }$ [see Eq.\ (\ref{MCurr}) of the
Appendix and (\ref{MCc}) below] will be seen to be on par with the relation (%
\ref{eq20}) of $X$ to the full $T$, and it will be referred to as the
nonpolar photoproduction amplitude.

We can write now several alternative expressions for the dressed current (%
\ref{eq28}):

\begin{eqnarray}
J^{\mu } &=&j_{0}^{\mu }+\left\langle F^{\mu }\right| G_{0}\left| \Gamma
\right\rangle +\left\langle F\right| \left( G_{0}+G_{0}XG_{0}\right) \left|
b^{\mu }\right\rangle  \nonumber \\
&=&j_{0}^{\mu }+\left\langle F^{\mu }\right| G_{0}\left| \Gamma
\right\rangle +\left\langle \Gamma \right| G_{0}\left| b^{\mu }\right\rangle
\nonumber \\
&=&j_{0}^{\mu }+\left\langle F^{\mu }\right| G_{0}\left| \Gamma
\right\rangle +\left\langle F\right| G_{0}\left| m^{\mu }\right\rangle 
\nonumber \\
&=&\,j_{0}^{\mu }+\left\langle F^{\mu }\right| G_{0}\left| \Gamma
\right\rangle +\left\langle \Gamma \right| G_{0}\left| F^{\mu }\right\rangle
+\left\langle \Gamma \right| G_{0}g^{\mu }G_{0}\left| \Gamma \right\rangle
+\left\langle \Gamma \right| G_{0}U^{\mu }G_{0}\left| \Gamma \right\rangle
\,\,,  \label{eq31}
\end{eqnarray}
with $j_{0}^{\mu }=Q_{N}Z\gamma ^{\mu }$ being the bare current.\footnote{%
$Z$ appears here because we choose to work with properly normalized vertices
[cf. remarks after Eq.\ (\ref{eq22})]. $Z$ could be absorbed into the vertices
at the expense of having to multiply the final result by $\sqrt{Z}$ for
every external nucleon leg. For vanishing interaction $U$ or for vanishing
photon momentum, $J^{\mu }$ just becomes $Q_{N}\gamma ^{\mu }$, of course.}
A structurally similar result can be obtained from the work of van Antwerpen 
and Afnan \cite{afnan} upon combining their equations (5.11) and (5.15) (see 
also discussion in Sec.\ VI). 
\begin{figure}
\centerline{\psfig{file=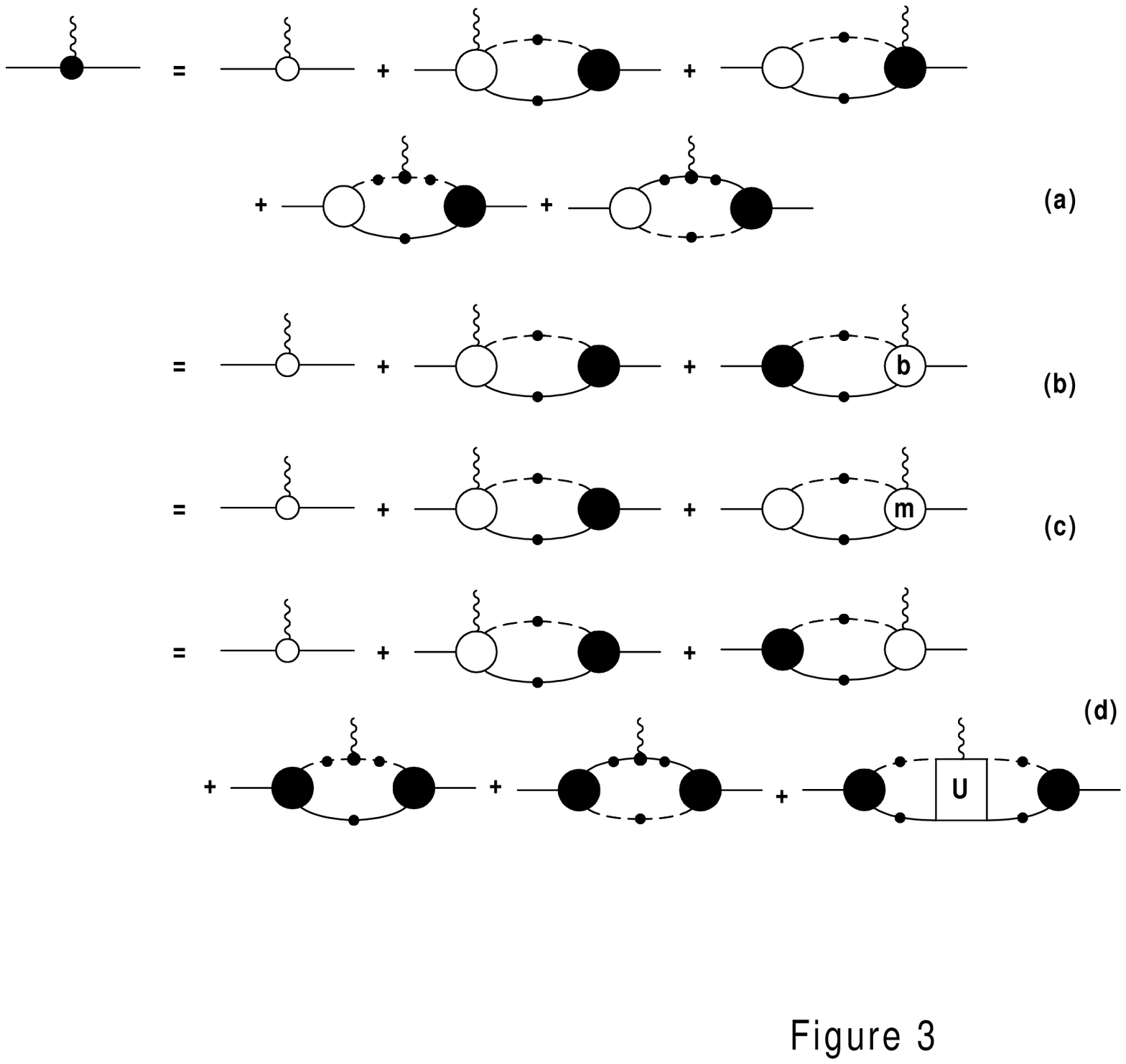,width=5in,clip=}}
\vspace{3mm}
\caption{Various equivalent representations of the fully dressed
electromagnetic current for the nucleon: (a) 
Eq.\ (\ref{eq28}); (b)-(d) correspond to
the last three equalities of 
Eq.\ (\ref{eq31}). Open circles depict bare and solid
ones fully dressed quantities. See also Figs.\ 4 and 6(a).}
\end{figure}

Figure 3 summarizes the various mechanisms of Eqs.\ (\ref{eq28}) and (\ref
{eq31}); Fig.\ 4 
shows Eq.\ (\ref{eq30-1}) and its driving term (\ref{eq30})
and Fig.\ 5 
depicts the interaction current $\Gamma ^{\mu }$ of the dressed
vertex, Eq.\ (\ref{eq29}). As can be read off these graphs, the description
of the dressed nucleon current given here is nonlinear, similar to what was
found already for the hadronic sector in Sec.\ II.
\begin{figure}[tbp]
\centerline{\psfig{file=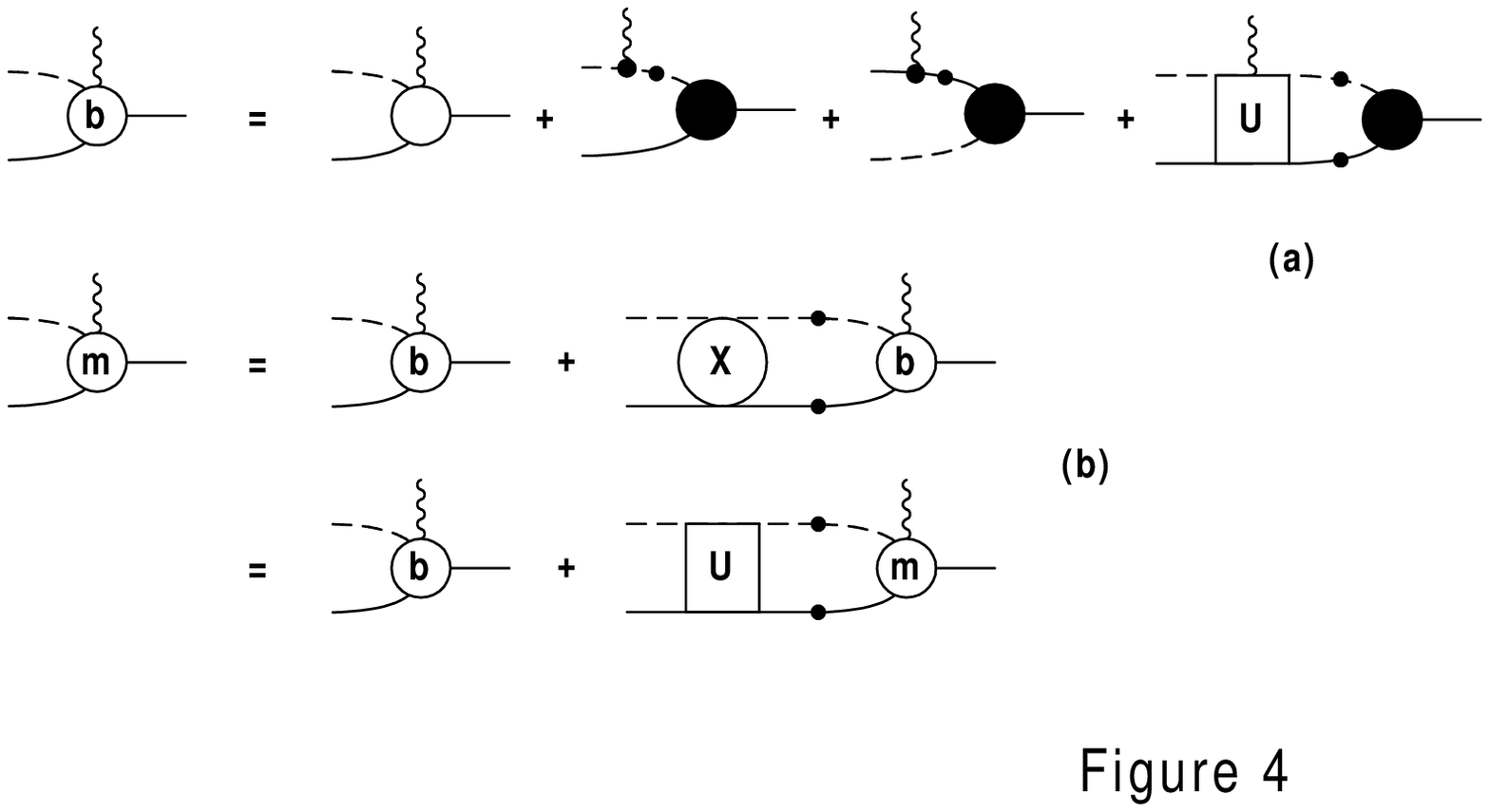,width=5in,clip=}}
\vspace{3mm}
\caption{(a) Born term of pion photoproduction amplitude, 
Eq.\ (\ref{eq30}). (b)
Auxiliary nonpolar amplitude, Eq.\ (\ref{eq30-1}).}
\end{figure}
\begin{figure} [btp]
\centerline{\psfig{file=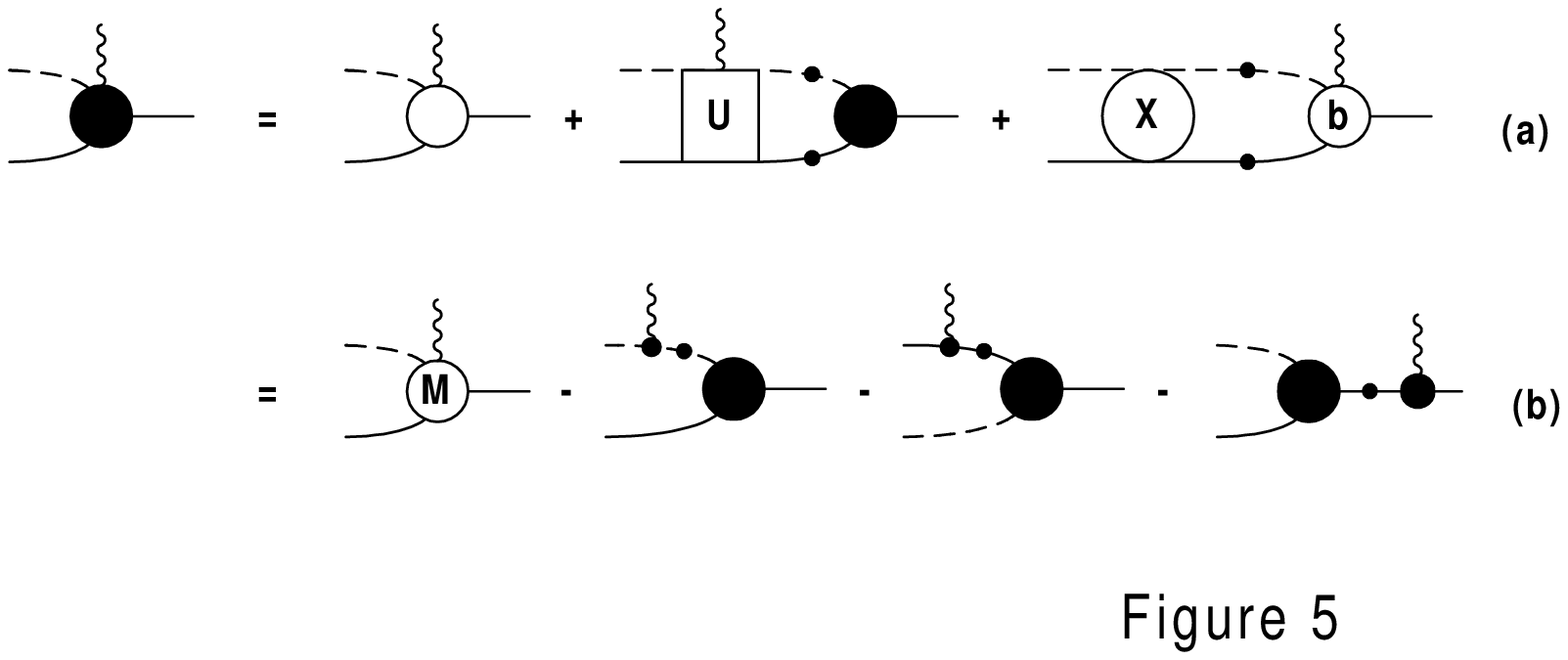,width=5in,clip=}}
\vspace{3mm}
\caption{Interaction current $\Gamma ^\mu$: (a) Definition, 
Eq.\ (\ref{eq29}), and (b)
relation to full pion photoproduction amplitude $M^\mu$, 
Eq.\ (\ref{ICurr}). }
\end{figure}

%%%%%%%%%%%%%%%%%%%%%%%%%%%%%%%%%%%%%%%%%%%%%%%%%%%%%%%%%%%%%%%%%%%%%%%%

\subsection{The exchange current $U^{\mu }$}

In order to exhibit the structure of the exchange current, 
for simplicity we concentrate only on the first two terms of
Eqs.\ (\ref{eq24}) and (\ref{eq25}) for the nucleon,

\begin{equation}
U^{\mu }=-\{U_{N}\}^{\mu }=-\{\left\langle \Gamma _{N}\right| \bar{S}\left|
\Gamma _{N}\right\rangle +\left\langle \Gamma _{N}\right| \left( \bar{S}X%
\bar{S}\right) \circ \Delta \left| \Gamma _{N}\right\rangle +...\}^{\mu
}\,\,\,.  \label{eq33}
\end{equation}
In the following we suppress the index $N$. We find

\begin{equation}
U^{\mu }=U_{0}^{\mu }+U_{1}^{\mu }+...  \label{eq33-1}
\end{equation}
with 
\begin{equation}
U_{0}^{\mu }=\left\langle \Gamma ^{\mu }\right| \overline{S}\left| \Gamma
\right\rangle +\left\langle \Gamma \right| \overline{SJ^{\mu }S}\left|
\Gamma \right\rangle +\left\langle \Gamma \right| \overline{S}\left| \Gamma
^{\mu }\right\rangle  \label{eq33-2}
\end{equation}
originating from the first graph in Fig.\ 2 and

\begin{eqnarray}
U_{1}^{\mu } &=&\left\langle \Gamma ^{\mu }\right| \left( \overline{S}X\,%
\overline{S}\right) \circ \Delta \left| \Gamma \right\rangle +\left\langle
\Gamma \right| \left( \overline{S}\,X\,\overline{S}\right) \circ \Delta
\left| \Gamma ^{\mu }\right\rangle +\left\langle \Gamma \right| \left( 
\overline{SJ^{\mu }S}\,\,X\,\overline{S}\right) \circ \Delta \left| \Gamma
\right\rangle \,  \nonumber \\
&&+\left\langle \Gamma \right| \left( \overline{S}\,X\,\overline{S}\right)
\circ \left( \Delta J_{\pi }^{\mu }\Delta \right) \left| \Gamma
\right\rangle +\left\langle \Gamma \right| \left( \overline{S}\,X^{\mu \,}%
\overline{S}\right) \circ \Delta \left| \Gamma \right\rangle +\left\langle
\Gamma \right| \left( \overline{S}\,X\overline{\,SJ^{\mu }S}\right) \circ
\Delta \left| \Gamma \right\rangle  \label{eq34}
\end{eqnarray}
from the second; the overbar again signifies exchanged hadrons. $X^{\mu }$
is given by

\begin{equation}
X^{\mu } =-\{X\}^{\mu } = (1+XG_{0})U^{\mu }(G_{0}X+1)+XG_{0}g^{\mu
}G_{0}X\,\,\,,  \label{eq35}
\end{equation}
as can be found from the gauge derivative of Eq.\ (\ref{eq18}). (The same
result was obtained in Eq. (5.8) of \cite{afnan}.) These
equations are summarized graphically in Fig.\ 6. 
\begin{figure}
\centerline{\psfig{file=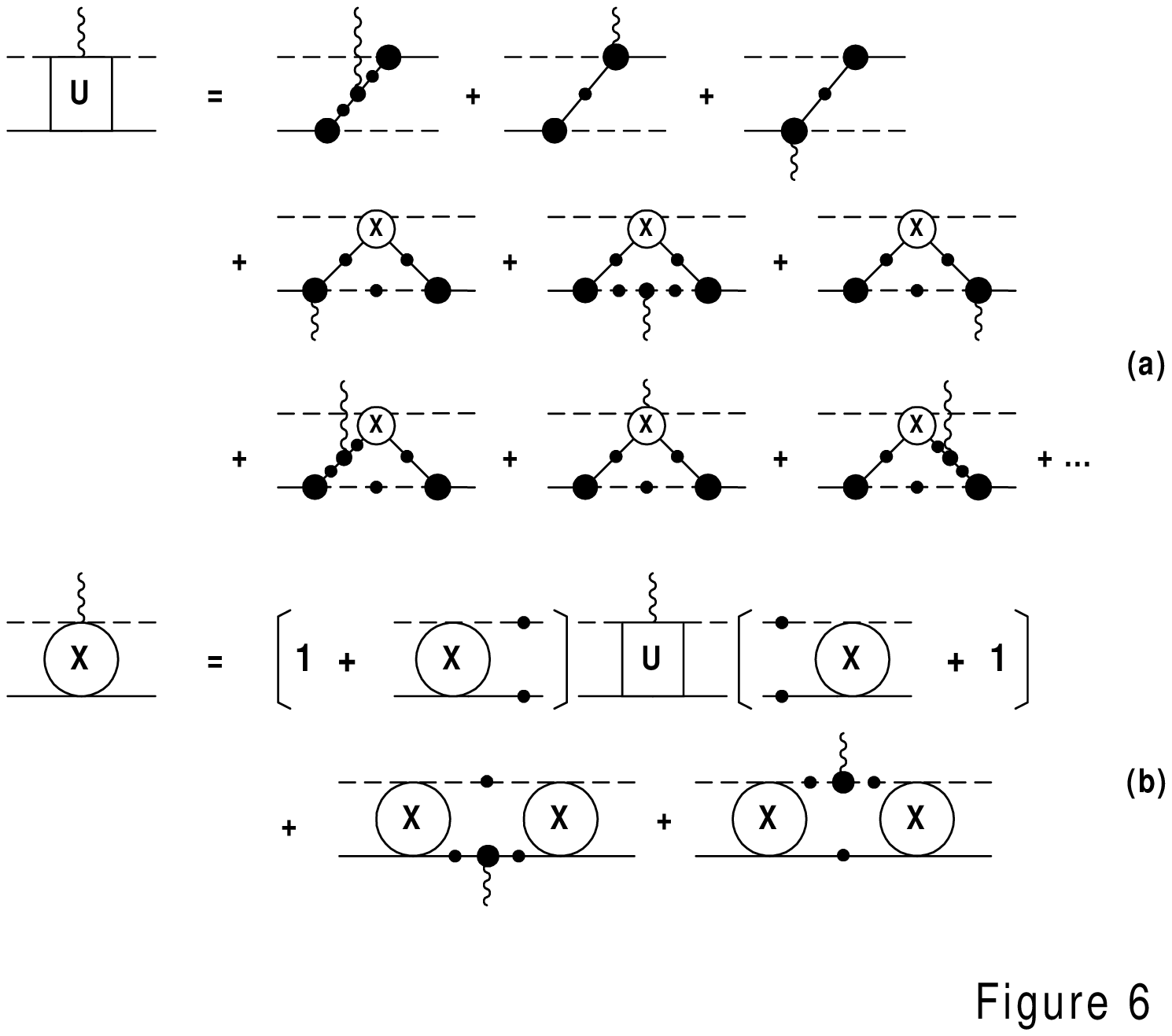,width=5in,clip=}}
\vspace{3mm}
\caption{(a) Exchange current $U^\mu$, 
Eq.\ (\ref{eq33}), and (b) current $X^\mu$,
Eq.\ (\ref{eq35}), 
associated with the photon attaching itself inside the hadronic
interaction region described by the nonpolar amplitude $X$.}
\end{figure}
In general, every vertex,
every internal propagator, and every transition amplitude generates a
separate contribution to the exchange current. Therefore, there are three
contributions to $U_{0}^{\mu }$ and six for $U_{1}^{\mu }$. As we shall
discuss below in the context of gauge invariance, when making approximations
mandated by practical considerations, one may omit entire pieces from the
exchange current $U^{\mu }$ without violating gauge invariance, as long as
one keeps together all pieces originating from the same hadron graph. For
example, omitting $U_{1}^{\mu }$ in its entirety would not violate gauge
invariance, but neglecting just one or two pieces in general would.

%%%%%%%%%%%%%%%%%%%%%%%%%%%%%%%%%%%%%%%%%%%%%%%%%%%%%%%%%%%%%%%%%%%%%%%%

\subsection{The pion production current}

The current $M^{\mu }(k;p^{\prime },p)$ for a photon of momentum $k$ hitting
a nucleon of initial momentum $p$ to produce a pion with momentum $%
q=p+k-p^{\prime },$ leaving behind the nucleon with momentum $p^{\prime }$,
according to Eq.\ (\ref{MCurr}) of the Appendix, is given by 
\begin{eqnarray}
M^{\mu }(k;p^{\prime },p) &=&-S^{-1}(p^{\prime })\Delta ^{-1}(p-p^{\prime
}+k)\,\{S(p^{\prime })\Delta (p-p^{\prime })\Gamma (p^{\prime
},p)S(p)\}^{\mu }S^{-1}(p)  \nonumber \\
&=&-\Gamma (p^{\prime },p+k)\{S(p)\}^{\mu }\,S^{-1}(p)  \nonumber \\
&&-\Delta ^{-1}(p-p^{\prime }+k)\{\Delta (p-p^{\prime })\}^{\mu }\Gamma
(p^{\prime },p)  \nonumber \\
&&-S^{-1}(p^{\prime })\{S(p^{\prime }-k)\}^{\mu }\,\Gamma (p^{\prime
}-k,p)\,\,\,  \nonumber \\
&&-\{\Gamma (p^{\prime },p)\}^{\mu }\,\,\,.  \label{MCdet}
\end{eqnarray}
\narrowtext
\noindent In a more concise notation which suppresses the momenta, the
preceding equation can be written as

\begin{eqnarray}
\left| M^{\mu }\right\rangle &=&-G_{0}^{-1}\,\{G_{0}\left| \Gamma
\right\rangle S\}^{\mu }\,S^{-1}  \nonumber \\
&=&g^{\mu }G_{0}\left| \Gamma \right\rangle +\left| \Gamma ^{\mu
}\right\rangle +\left| \Gamma \right\rangle \,S\,J^{\mu }  \nonumber \\
&=&\left| \Gamma \right\rangle \,S\,J^{\mu }+(1+XG_{0})\left| b^{\mu
}\right\rangle  \nonumber \\
&=&\left| \Gamma \right\rangle \,S\,J^{\mu }+\left| m^{\mu }\right\rangle
\,\,\,,  \label{MCc}
\end{eqnarray}
where use was made of Eqs.\ (\ref{eq29})-(\ref{eq30-1}) to simplify the
result. The only difference, therefore, of the full current $M^{\mu }$ and
the nonpolar current $m^{\mu }$ of Eq.\ (\ref{eq30-1}) is the pole term $%
\left| \Gamma \right\rangle \,S\,J^{\mu }$, where the reaction proceeds
through an intermediary nucleon propagator (see Fig.\ 7).
\begin{figure}
\centerline{\psfig{file=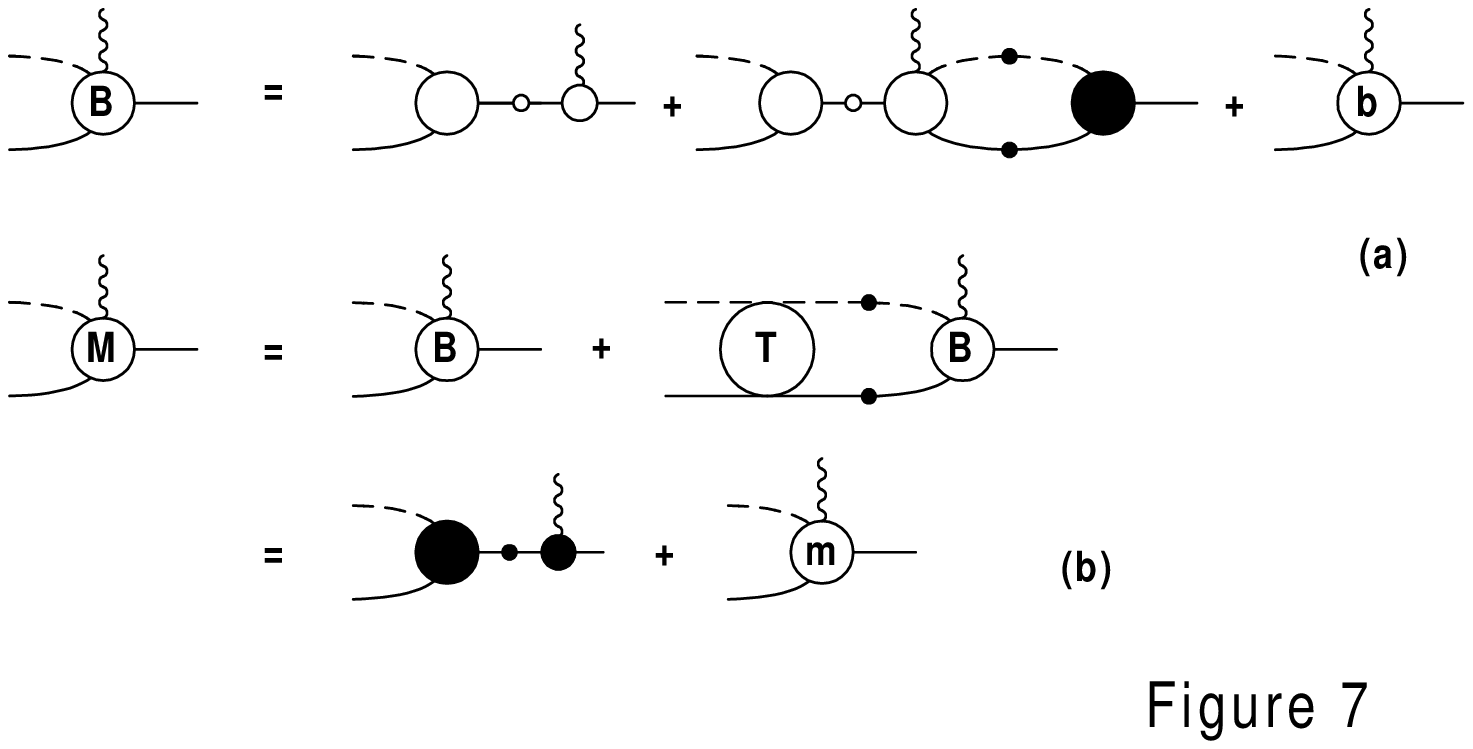,width=5in,clip=}}
\vspace{3mm}
\caption{(a) Alternative Born term $B^\mu$, 
Eq.\ (\ref{BT}), for the full pion
photoproduction amplitude $M^\mu$ if the $\pi N$ final-state interaction is
described by the full $T$ instead of by the nonpolar $X$; (b) shows the
resulting two, completely equivalent describtions of $M^\mu$, 
Eqs.\ (\ref{MT}) and
(\ref{MCc}) [see also Fig.\ 4(b)].}
\end{figure}

The vertex current proper, i.e., that piece of the production amplitude
where the photon attaches itself within the hadronic interaction region, and
not to anyone of the three legs of the $\pi NN$ vertex, which is described
by $\Gamma ^{\mu }$ of Eq.\ (\ref{eq29}), is then given by 
\begin{equation}
\left| \Gamma ^{\mu }\right\rangle =\left| M^{\mu }\right\rangle -g^{\mu
}G_{0}\left| \Gamma \right\rangle -\left| \Gamma \right\rangle \,S\,J^{\mu
}\,\,\,.  \label{ICurr}
\end{equation}
This well-known result \cite{kazes,friar,afnan} is sometimes referred to
as a ``contact term.'' In view of its rich internal dynamical structure
exhibited in Fig.\ 5, we prefer to call it an interaction current. Indeed, at
the level of fully dressed, physical particles, it is the only term that
contains any final-state interaction between pions and nucleons.

As an addendum to the last remark, we mention that the fully dressed pole
term $\left| \Gamma \right\rangle \,S\,J^{\mu }$ may be viewed as coming
about through the final-state interaction mediated by the full $T$, of Eq.\ (%
\ref{eq16}), rather than by the nonpolar $X$, as in Eq.\ (\ref{eq30-1}). To
this end, we mention without derivation that, instead of via $m^{\mu }$, as
in (\ref{MCc}), $M^{\mu }$ can be obtained directly as

\begin{equation}
\left| M^{\mu }\right\rangle =\left| B^{\mu }\right\rangle +TG_{0}\left|
B^{\mu }\right\rangle \,\,\,,  \label{MT}
\end{equation}
where the modified Born term $B^{\mu }$ is now

\begin{equation}
\left| B^{\mu }\right\rangle =\left| F\right\rangle S_{0}J_{0}^{\mu }+\left|
F\right\rangle S_{0}\left\langle F^{\mu }\right| G_{0}\left| \Gamma
\right\rangle +\left| b^{\mu }\right\rangle \,\,\,,  \label{BT}
\end{equation}
in other words, as compared to $b^{\mu }$ of Eq.\ (\ref{eq30}), it contains
explicit bare hadronic vertices and propagators. It is the final-state
interaction in $T$ which produces the direct dressed pole term $\left|
\Gamma \right\rangle \,S\,J^{\mu }$ and reduces the remaining final-state
interaction to one via $X$. These relations are also depicted in Fig.\ 7.

%%%%%%%%%%%%%%%%%%%%%%%%%%%%%%%%%%%%%%%%%%%%%%%%%%%%%%%%%%%%%%%%%%%%%%%%

\section{Gauge invariance}

The gauge invariance of the electromagnetic interaction requires 
\cite{weinberg} that the divergencies of all physical currents vanish if all
external hadrons are on their respective mass shells. In technical terms,
this applies to all currents that are based on the reduction of an $n$-point
Green's function (see Appendix), i.e., to the pion current $J_{\pi }^{\mu }$%
, the nucleon current $J^{\mu }$ and the production current $M^{\mu }$:

\begin{mathletters}
\label{kj0}
\begin{eqnarray}
k_{\mu }J_{\pi }^{\mu }(q+k,q) &=&0\,\,\,,  \label{gjpi} \\
k_{\mu }J^{\mu }(p+k,p) &=&0\,\,\,,  \label{gj} \\
k_{\mu }M^{\mu }(k;p^{\prime },p) &=&0\,\,\,,  \label{gm}
\end{eqnarray}
\end{mathletters}
It does not apply, for example, to the exchange current $U^{\mu }$ and the
interaction current $\Gamma ^{\mu }$ since they do not involve
electromagnetic interactions with external legs and therefore are not
directly observable. For these currents different gauge conditions apply, as
discussed below.

The key here are the Ward-Takahashi identities \cite{WTI} for the off-shell
propagators,

\begin{mathletters}
\label{WT}
\begin{eqnarray}
k_{\mu }J_{\pi }^{\mu }(q^{\prime },q) &=&\Delta ^{-1}(q^{\prime })Q_{\pi
}-Q_{\pi }\Delta ^{-1}(q)\,\,\,,  \label{WTpi} \\
k_{\mu }J^{\mu }(p^{\prime },p) &=&S^{-1}(p^{\prime
})Q_{N}-Q_{N}S^{-1}(p)\,\,\,,  \label{WTN}
\end{eqnarray}
\end{mathletters}
where

\begin{eqnarray}
Q_{\pi ,ij} &=&ie\varepsilon _{i3j}\,\,\,, \\
Q_{N} &=&\frac{e}{2}(1+\tau _{3})\,\,\,,
\end{eqnarray}
are the respective charge operators. Note that the placement of the charge
operators in Eqs.\ (\ref{WT}) is mindful of the fact that for dressed
particles, the self-energies within the propagators carry isospin dependence
and therefore do not commute {\it a priori} with the charge operators. One
may argue, of course, that this subtlety is largely academic since the
Ward-Takahashi identity is a statement about charge conservation and
therefore immediately implies $[\Sigma ,Q]=0$. Nevertheless, since in the
present formulation the placements of the charge operators will carry the
additional information where, and to which particle, the photon momentum is
fed into the equations, we will strictly apply the Ward-Takahashi identities
in the form given here.

Taking the divergence of the production current $M^{\mu }$ given by Eq.\ (\ref
{MCc}) and assuming the validity of the Ward-Takahashi identities readily
gives \cite{kazes,afnan} \widetext
\begin{eqnarray}
k_{\mu }\left| M_{p+k,p}^{\mu }\tau \right\rangle &=&\Delta
^{-1}(p-p^{\prime }+k)\widehat{Q}_{\pi }\Delta (p-p^{^{\prime }})\left|
\Gamma _{p}\tau \right\rangle +S^{-1}(p^{\prime })\widehat{Q}_{N}S(p^{\prime
}-k)\left| \Gamma _{p}\tau \right\rangle  \nonumber \\
&&-\left| \Gamma _{p+k}\tau \right\rangle S(p+k)\widehat{Q}%
_{N}S^{-1}(p)\,\,\,  \nonumber \\
&&+k_{\mu }\left| \Gamma _{p+k,p}^{\mu }\tau\right\rangle +\left| \Gamma
_{p+k}\tau \right\rangle \widehat{Q}_{N}-\widehat{Q}_{\pi }\left| \Gamma
_{p}\tau \right\rangle -\widehat{Q}_{N}\left| \Gamma _{p}\tau \right\rangle
\,\,\,,  \label{MG1}
\end{eqnarray}
\narrowtext
\noindent where the momentum indices exhibit the total available {\it %
hadronic} momentum and $p^{\prime }$ is the momentum of the final nucleon.
Entities carrying a photon index $\mu $ have two indices since the hadron
momentum available after the electromagnetic interaction is increased by the
photon's momentum $k$. We have also now explicitly included the symbol $\tau 
$ in the kets as a mnemonic that there is an isospin index associated with
each vertex and to remind us that one cannot simply commute charge operators
with vertices. From the context it will be clear how to choose this index.

In the notation adopted in (\ref{MG1}) the meaning of the charge operator $%
Q_{x}$ of particle $x$ has now been expanded: In addition to performing the
usual charge operation $Q_{x}$, $\widehat{Q}_{x}$ adds a photon momentum $k$
to the {\it charged} particle $x$ at the corresponding place in the
equations, e.g., $\widehat{Q}_{\pi }\Delta (p-p^{^{\prime }})\left| \Gamma
_{p}\tau \right\rangle $ means that for all subsequent interactions the pion
coming out of the vertex with momentum $p-p^{\prime }$ will have momentum $%
p-p^{\prime }+k$. With all external momenta fixed, the placement and
particle index of $Q_{x}$ allows one to unambiguously determine all internal
momenta (except for loop integrations, of course) at every stage of the
reaction. For example, in $\left| \Gamma _{p+k}\tau \right\rangle \widehat{Q}%
_{N} $, we could drop the total-momentum index $p+k$ since the rule tells us
that $\widehat{Q}_{N}$ will add a photon momentum $k$ to the incoming
momentum $p $ to provide an initial nucleon momentum $p+k$ for the vertex $%
\left| \Gamma \tau \right\rangle .$ This short-hand notation will turn out
to be extremely useful in keeping the following expressions as concise as
possible without becoming sketchy.

Since the first three terms on the right-hand side of Eq.\ (\ref{MG1}) vanish
on-shell, the current conservation is tantamount to the well-known \cite
{kazes,afnan} condition 
\begin{equation}
k_{\mu }\left| \Gamma _{p+k,p}^{\mu }\tau \right\rangle +\left| \Gamma
_{p+k}\tau \right\rangle \widehat{Q}_{N}-\widehat{Q}_{\pi }\left| \Gamma
_{p}\tau \right\rangle -\widehat{Q}_{N}\left| \Gamma _{p}\tau \right\rangle
=0  \label{Gcont}
\end{equation}
for the interaction current $\Gamma ^{\mu }$. Note that the form of this
condition is similar to a continuity equation with a surface term,

\begin{equation}
k_{\mu }\Gamma ^{\mu }+\Delta _{k}R_{\Gamma }=0\,\,\,,  \label{Gcont1}
\end{equation}
where the ``surface term''

\begin{equation}
\Delta _{k}R_{\Gamma }=\left| \Gamma _{p+k}\tau \right\rangle \widehat{Q}%
_{N}-\widehat{Q}_{\pi }\left| \Gamma _{p}\tau \right\rangle -\widehat{Q}%
_{N}\left| \Gamma _{p}\tau \right\rangle \,\,\,  \label{rhoG}
\end{equation}
measures the sum of all changes in the internal reaction dynamics brought
about when a photon momentum $k$ is transmitted through the hadronic system
from an incoming charged particle to an outgoing charged particle; of
course, within the interaction region $k$ can be shared with any particle,
charged or uncharged. This quantity is illustrated in Fig.\ 8 
\begin{figure}
\centerline{\psfig{file=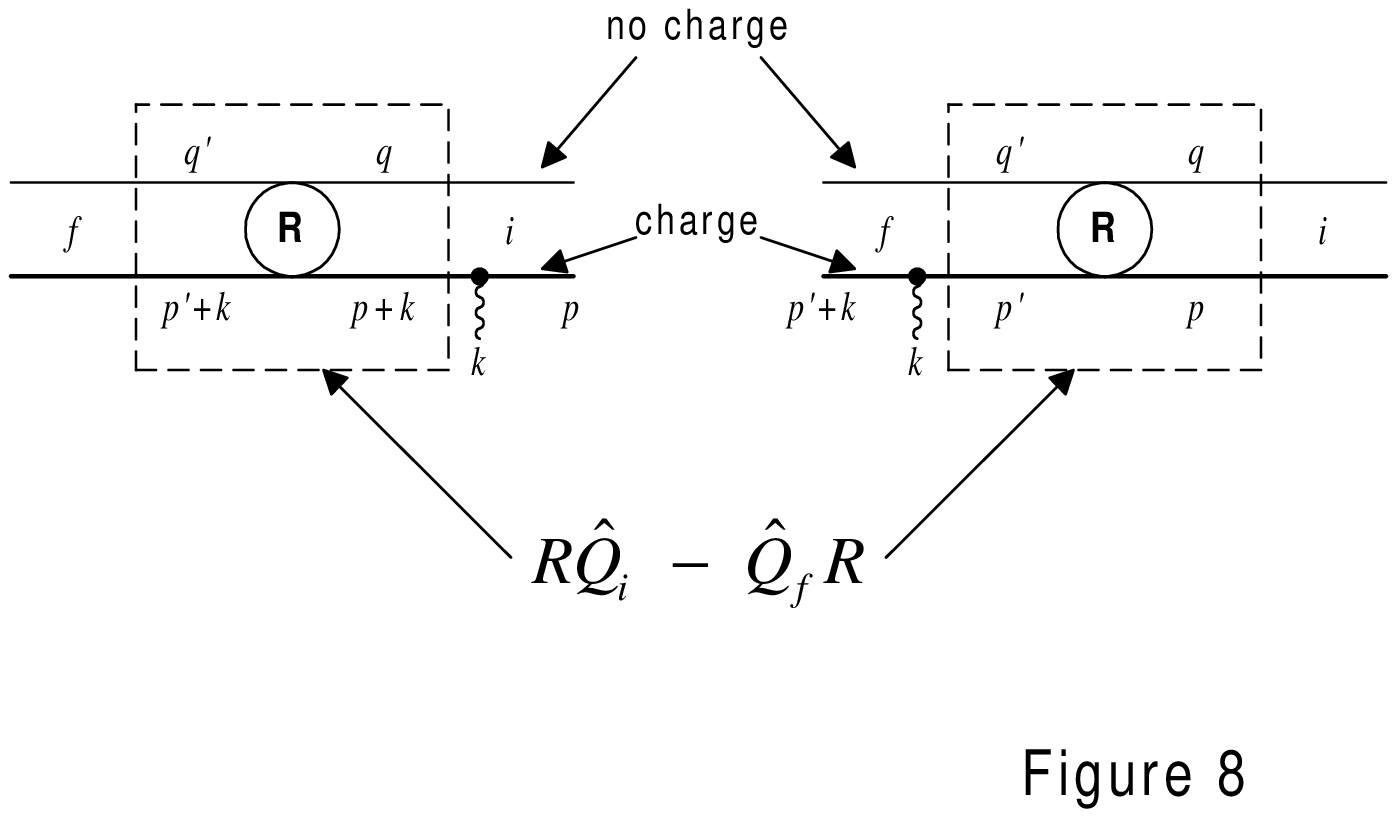,width=5in,clip=}}
\vspace{3mm}
\caption{Generic representation of the photonic reaction change 
[cf. Eq.\ (\ref{Ttilde})]. 
$R$ is an arbitrary hadronic reaction mechanism where all incoming
and outgoing {\it uncharged} particles have been subsumed in the upper lines
and all incoming and outgoing {\it charged} particles in the lower, thicker
lines. The first graph on the left sums up all contributions where the
photon is attached to an incoming particle whereas the one on the right
depicts the sum of all contributions from the photon being attached to an
outgoing particle. The photonic reaction change is the difference between
the purely hadronic contributions enclosed in the dashed boxes; it measures
the change brought about in the hadronic reaction when a photon momentum is
transmitted through the hadronic interaction region entering and leaving via
charged particles.}
\end{figure}
for a somewhat
more general case. For brevity, we will refer to $\Delta _{k}R_{\Gamma }$ as
the (purely hadronic) photonic reaction change or simply reaction change. As
will be shown presently, continuity equations similar to (\ref{Gcont1}),
with $\Delta _{k}R$'s exactly analogous to Eq.\ (\ref{rhoG}), govern all
aspects of the interaction of photons with hadrons.

In the following, we will prove gauge invariance of the formalism developed
here by showing that the Ward-Takahashi identity (\ref{WTN}) and the
continuity equation (\ref{Gcont}) hold true. In view of the nonlinearities
of the present equations, a direct proof does not seem possible and the
proof will be one of self-consistency. In other words, we first show that
assuming the validity of the Ward-Takahashi identities leads to the
continuity equation for the interaction current and then, second, proceed
that we can verify the self-consistency of the assumption by deriving (\ref
{WTN}) using the details of $J^{\mu }$ given by Eq.\ (\ref{eq28}).

%%%%%%%%%%%%%%%%%%%%%%%%%%%%%%%%%%%%%%%%%%%%%%%%%%%%%%%%%%%%%%%%%%%%%%%%

\subsection{The gauge condition for the interaction current}

With Eq.\ (\ref{eq29}) the divergence of the interaction current $\Gamma
^{\mu }$ is written as

\begin{equation}
k_{\mu }\left| \Gamma ^{\mu }\right\rangle =k_{\mu }\left( \left| F^{\mu
}\right\rangle +U^{\mu }G_{0}\left| \Gamma \right\rangle +XG_{0}\left|
b^{\mu }\right\rangle \right) \,\,\,.
\end{equation}
In a more detailed notation, where isospin and the dependence on the
corresponding total hadron momenta is shown, we have \widetext
\begin{equation}
k_{\mu }\left| \Gamma _{p+k,p}^{\mu }\tau \right\rangle =k_{\mu }\left(
\left| F_{p+k,p}^{\mu }\tau \right\rangle +U_{p+k,p}^{\mu }G_{0,p}\left|
\Gamma _{p}\tau \right\rangle +X_{p+k}G_{0,p+k}\left| b_{p+k,p}^{\mu }\tau
\right\rangle \right) \,\,\,.  \label{kG}
\end{equation}
Now, as a first step, let us define

\begin{equation}
\widetilde{U}=k_{\mu }U_{p+k,p}^{\mu }+U_{p+k}\widehat{Q}_{\pi }+U_{p+k}%
\widehat{Q}_{N}-\widehat{Q}_{\pi }U_{p}-\widehat{Q}_{N}U_{p}\,\,\,\,.
\label{Wdef}
\end{equation}
The right-hand side here is seen to be constructed in exact analogy to the
left-hand side of the continuity equation (\ref{Gcont}) with

\begin{equation}
\Delta _{k}R_{U}=U_{p+k}\widehat{Q}_{\pi }+U_{p+k}\widehat{Q}_{N}-\widehat{Q}%
_{\pi }U_{p}-\widehat{Q}_{N}U_{p}\,\,
\end{equation}
being the corresponding photonic reaction change. We conjecture, therefore,
that (\ref{Wdef}) {\it is} indeed a continuity equation and that 
\begin{equation}
\widetilde{U}=0\,\,\,;  \label{Utilde0}
\end{equation}
hence, 
\begin{equation}
k_{\mu }U_{p+k,p}^{\mu }+\Delta _{k}R_{U}=0\,\,\,.  \label{Wdef0}
\end{equation}
In the subsection IV.B, we will show that this conjecture is indeed valid.

Using 
\begin{equation}
\widehat{Q}_{x}UG_{0}\left| \Gamma \tau \right\rangle =\,\widehat{Q}%
_{x}\left| \Gamma \tau \right\rangle -\widehat{Q}_{x}\left| F\tau
\right\rangle \,\,\,,
\end{equation}
we then have for (\ref{kG}),

\begin{eqnarray}
k_{\mu }\left| \Gamma _{p+k,p}^{\mu }\tau \right\rangle &=&k_{\mu }\left|
F_{p+k,p}^{\mu }\tau \right\rangle -\widehat{Q}_{\pi }\left| F_{p}\tau
\right\rangle -\widehat{Q}_{N}\left| F_{p}\tau \right\rangle  \nonumber \\
&&+\widehat{Q}_{\pi }\left| \Gamma _{p}\tau \right\rangle +\widehat{Q}%
_{N}\left| \Gamma _{p}\tau \right\rangle  \nonumber \\
&&-\left( U_{p+k}\widehat{Q}_{\pi }+U_{p+k}\widehat{Q}_{N}\right)
G_{0,p}\left| \Gamma _{p}\tau \right\rangle  \nonumber \\
&&+X_{p+k}G_{0,p+k}k_{\mu }\left| b_{p+k,p}^{\mu }\tau \right\rangle
\,\,\,\,\,.  \label{kG1}
\end{eqnarray}
To simplify this further, let us look at $k_{\mu }b^{\mu }$, using Eqs.\ (\ref
{eq30}) and (\ref{Wdef0}) and the Ward-Takahashi identities (\ref{WT}),

\begin{eqnarray}
k_{\mu }\left| b_{p+k,p}^{\mu }\tau \right\rangle &=&k_{\mu }\left|
F_{p+k,p}^{\mu }\tau \right\rangle -\widehat{Q}_{\pi }\left| F_{p}\tau
\right\rangle -\widehat{Q}_{N}\left| F_{p}\tau \right\rangle +\widehat{Q}%
_{\pi }\left| \Gamma _{p}\tau \right\rangle +\widehat{Q}_{N}\left| \Gamma
_{p}\tau \right\rangle  \nonumber \\
&&+\left[ \Delta ^{-1}(p-p^{\prime }+k)\widehat{Q}_{\pi }-\widehat{Q}_{\pi
}\Delta ^{-1}(p-p^{\prime })\right] \Delta (p-p^{\prime })\left| \Gamma
_{p}\tau \right\rangle  \nonumber \\
&&+\left[ S^{-1}(p^{\prime }+k)\widehat{Q}_{N}-\widehat{Q}%
_{N}S^{-1}(p^{\prime })\right] S(p^{\prime })\left| \Gamma _{p}\tau
\right\rangle  \nonumber \\
&&-\left( U_{p+k}\widehat{Q}_{\pi }+U_{p+k}\widehat{Q}_{N}\right)
G_{0,p}\left| \Gamma _{p}\tau \right\rangle  \nonumber \\
&=&k_{\mu }\left| F_{p+k,p}^{\mu }\tau \right\rangle -\widehat{Q}_{\pi
}\left| F_{p}\tau \right\rangle -\widehat{Q}_{N}\left| F_{p}\tau
\right\rangle  \nonumber \\
&&+\Delta ^{-1}(p-p^{\prime }+k)\widehat{Q}_{\pi }\Delta (p-p^{\prime
})\left| \Gamma _{p}\tau \right\rangle  \nonumber \\
&&+S^{-1}(p^{\prime }+k)\widehat{Q}_{N}S(p^{\prime })\left| \Gamma _{p}\tau
\right\rangle  \nonumber \\
&&-\left( U_{p+k}\widehat{Q}_{\pi }+U_{p+k}\widehat{Q}_{N}\right)
G_{0,p}\left| \Gamma _{p}\tau \right\rangle \,\,\,,  \label{kb1}
\end{eqnarray}
where $p^{\prime }$ is the momentum of the outgoing nucleon.

Introducing now a continuity-equation term in analogy to (\ref{Wdef}), 
\begin{equation}
\left| \widetilde{F}\right\rangle =k_{\mu }\left| F_{p+k,p}^{\mu }\tau
\right\rangle +\left| F_{p+k}\tau \right\rangle \widehat{Q}_{N}-\widehat{Q}%
_{\pi }\left| F_{p}\tau \right\rangle -\widehat{Q}_{N}\left| F_{p}\tau
\right\rangle \,\,\,\,,  \label{Gdef}
\end{equation}
and collecting all partial results, Eq.\ (\ref{kG1}) becomes

\begin{eqnarray}
k_{\mu }\left| \Gamma _{p+k,p}^{\mu }\tau \right\rangle &=&\left| \widetilde{%
F} \right\rangle +X_{p+k}G_{0,p+k}\left| \widetilde{F}%
\right\rangle  \nonumber \\
&&+\widehat{Q}_{\pi }\left| \Gamma _{p}\tau \right\rangle +\widehat{Q}%
_{N}\left| \Gamma _{p}\tau \right\rangle  \nonumber \\
&&-\left| F_{p+k}\tau \right\rangle \widehat{Q}_{N}-X_{p+k}G_{0,p+k}\left|
F_{p+k}\tau \right\rangle \widehat{Q}_{N}  \nonumber \\
&&-\left( U_{p+k}\widehat{Q}_{\pi }+U_{p+k}\widehat{Q}_{N}\right)
G_{0,p}\left| \Gamma _{p}\tau \right\rangle  \nonumber \\
&&+\left( X_{p+k}\widehat{Q}_{\pi }+X_{p+k}\widehat{Q}_{N}\right)
G_{0,p}\left| \Gamma _{p}\tau \right\rangle  \nonumber \\
&&-\left( X_{p+k}G_{0,p+k}U_{p+k}\widehat{Q}_{\pi }+X_{p+k}G_{0,p+k}U_{p+k}%
\widehat{Q}_{N}\right) G_{0,p}\left| \Gamma _{p}\tau \right\rangle \,\,\,,
\label{kGam}
\end{eqnarray}
where the last three terms cancel and one finally obtains 
\begin{equation}
k_{\mu }\left| \Gamma _{p+k,p}^{\mu }\tau \right\rangle +\left| \Gamma
_{p+k}\tau \right\rangle \widehat{Q}_{N}-\widehat{Q}_{\pi }\left| \Gamma
_{p}\tau \right\rangle -\widehat{Q}_{N}\left| \Gamma _{p}\tau \right\rangle
=\left( 1+X_{p+k}G_{0,p+k}\right) \left| \widetilde{F}\right\rangle
\,\,\,.\,
\end{equation}
We thus find that a sufficient condition for the validity of the continuity
equation (\ref{Gcont}) is that $\widetilde{F}=0$, in other words, that (\ref
{Gdef}) become a proper continuity equation itself.

At this stage, in view of its ambiguity in an effective field theory based
on hadronic degrees of freedom only, i.e., without any reaction-dynamical
basis for discerning mechanisms that contribute to the bare contact current $%
F^{\mu }$, we may simple demand that $F^{\mu }$ be such that 
\begin{equation}
\left| \widetilde{F}\right\rangle =0\,\,\,.  \label{Gzero}
\end{equation}
It then follows that the interaction current satisfies the continuity
equation (\ref{Gcont}) and hence the divergence of the production current,

\begin{eqnarray}
k_{\mu }\left| M_{p+k,p}^{\mu }\tau \right\rangle &=&\Delta
^{-1}(p-p^{\prime }+k)\widehat{Q}_{\pi }\Delta (p-p^{^{\prime }})\left|
\Gamma _{p}\tau \right\rangle +S^{-1}(p^{\prime })\widehat{Q}_{N}S(p^{\prime
}-k)\left| \Gamma _{p}\tau \right\rangle  \nonumber \\
&&-\left| \Gamma _{p+k}\tau \right\rangle S(p+k)\widehat{Q}%
_{N}S^{-1}(p)\,\,\,\,,
\end{eqnarray}
vanishes on-shell, thus making $M^{\mu }$ gauge invariant.

In Sec.\ V, a physical justification for the requirement (\ref{Gzero}) will
be given.

%%%%%%%%%%%%%%%%%%%%%%%%%%%%%%%%%%%%%%%%%%%%%%%%%%%%%%%%%%%%%%%%%%%%%%%%

\subsection{The gauge condition for the exchange current}

We need to verify now that the gauge condition (\ref{Utilde0}) is indeed
satisfied for $U^{\mu }$. In order to do so, we will make liberal use of the
facility offered by the $\widehat{Q}_{x}$ notation to keep track of where a
photon momentum needs to be injected into the equation. Note that the
Ward-Takahashi identities (\ref{WT}) can be written as simple commutators,

\begin{mathletters}
\label{qcNp}
\begin{eqnarray}
k_{\mu }SJ^{\mu }S &=&\widehat{Q}_{N}S-S\widehat{Q}_{N}\,\,\,\,,  \label{qcN}
\\
k_{\mu }\Delta J_{\pi }^{\mu }\Delta &=&\widehat{Q}_{\pi }\Delta -\Delta 
\widehat{Q}_{\pi }\,\,\,,  \label{qcpi}
\end{eqnarray}
\end{mathletters}
without any need for momentum arguments. Similarly, we have for the
continuity equation (\ref{Gcont}),

\begin{equation}
k_{\mu }\left| \Gamma ^{\mu }\tau \right\rangle =(\widehat{Q}_{\pi _{f}}+%
\widehat{Q}_{N_{f}})\left| \Gamma \tau \right\rangle -\left| \Gamma \tau
\right\rangle \widehat{Q}_{N_{i}}\,\,,
\end{equation}
where the indices $i$ and $f$ refer to initial and final, respectively. With
external momenta fixed, these short-hand versions allow one to determine
every momentum unambiguously. With this notation, what needs to be proved
now is [cf. Eq.\ (\ref{Wdef0})]

\begin{equation}
k_{\mu }U^{\mu }+U(\widehat{Q}_{N_{i}}+\widehat{Q}_{\pi _{i}})-(\widehat{Q}%
_{N_{f}}+\widehat{Q}_{\pi _{f}})U=0\,\,\,.  \label{Ucont3}
\end{equation}
To this end, we first consider the three terms of $U_{0}^{\mu }$ of Eq.\ (\ref
{eq33-2}), making use of the commutator notation just presented and
consulting Fig.\ 6(a):

\begin{eqnarray}
k_{\mu }U_{0}^{\mu } &=&k_{\mu }\left( \left\langle \Gamma ^{\mu }\right| 
\overline{S}\left| \Gamma \right\rangle +\left\langle \Gamma \right| 
\overline{SJ^{\mu }S}\left| \Gamma \right\rangle +\left\langle \Gamma
\right| \overline{S}\left| \Gamma ^{\mu }\right\rangle \right)  \nonumber \\
&=&\left[ \widehat{Q}_{N_{f}}\left\langle \Gamma \right| -\left\langle
\Gamma \right| (\widehat{Q}_{N}+\widehat{Q}_{\pi _{i}})\right] \overline{S}%
\left| \Gamma \right\rangle  \nonumber \\
&&+\left\langle \Gamma \right| \left( \widehat{Q}_{N}\overline{S}-\overline{S%
}\widehat{Q}_{N}\right) \left| \Gamma \right\rangle +\left\langle \Gamma
\right| \overline{S}\left[ (\widehat{Q}_{N}+\widehat{Q}_{\pi _{f}})\left|
\Gamma \right\rangle -\left| \Gamma \right\rangle \widehat{Q}_{N_{i}}\right]
\nonumber \\
&=&\widehat{Q}_{N_{f}}\left\langle \Gamma \right| \overline{S}\left| \Gamma
\right\rangle -\left\langle \Gamma \right| \widehat{Q}_{N}\overline{S}\left|
\Gamma \right\rangle -\left\langle \Gamma \right| \widehat{Q}_{\pi _{i}}%
\overline{S}\left| \Gamma \right\rangle -\left\langle \Gamma \right| 
\overline{S}\widehat{Q}_{N}\left| \Gamma \right\rangle  \nonumber \\
&&+\left\langle \Gamma \right| \widehat{Q}_{N}\overline{S}\left| \Gamma
\right\rangle +\left\langle \Gamma \right| \overline{S}\widehat{Q}_{N}\left|
\Gamma \right\rangle +\left\langle \Gamma \right| \overline{S}\widehat{Q}%
_{\pi _{f}}\left| \Gamma \right\rangle -\left\langle \Gamma \right| 
\overline{S}\left| \Gamma \right\rangle \widehat{Q}_{N_{i}}  \nonumber \\
&=&(\widehat{Q}_{N_{f}}+\widehat{Q}_{\pi _{f}})\left\langle \Gamma \right| 
\overline{S}\left| \Gamma \right\rangle -\left\langle \Gamma \right| 
\overline{S}\left| \Gamma \right\rangle (\widehat{Q}_{N_{i}}+\widehat{Q}%
_{\pi _{i}})  \nonumber \\
&=&(\widehat{Q}_{N_{f}}+\widehat{Q}_{\pi _{f}})U_{0}-U_{0}(\widehat{Q}%
_{N_{i}}+\widehat{Q}_{\pi _{i}})\,\,\,  \label{kU0}
\end{eqnarray}
which is the desired result ($\widehat{Q}_{N}$ here is the charge operator
of the exchanged nucleon). $U_{0}$ by itself, therefore, satisfies already
the continuity equation.

The key for calculating the divergence of $U_{1}^{\mu }$ is the divergence
of $X^{\mu }$ of Eq.\ (\ref{eq35}) since this is the only new piece [see also
Fig.\ 6(b)] required in the calculation of $k_{\mu }U_{1}^{\mu }$. One has

\begin{eqnarray}
k_{\mu }X^{\mu } &=&k_{\mu }(1+XG_{0})U^{\mu }(XG_{0}+1)+k_{\mu
}XG_{0}g^{\mu }G_{0}X\,  \nonumber \\
&=&(1+XG_{0})\left[ (\widehat{Q}_{N_{f}}^{i}+\widehat{Q}_{\pi _{f}}^{i})U-U(%
\widehat{Q}_{N_{i}}^{i}+\widehat{Q}_{\pi _{i}}^{i})\right] (XG_{0}+1) 
\nonumber \\
&&+X\left[ (\widehat{Q}_{N_{f}}^{i}+\widehat{Q}_{\pi _{f}}^{i})G_{0}-G_{0}(%
\widehat{Q}_{N_{i}}^{i}+\widehat{Q}_{\pi _{i}}^{i})\right] X\,\,\,  \nonumber
\\
&=&(1+XG_{0})(\widehat{Q}_{N_{f}}^{i}+\widehat{Q}_{\pi _{f}}^{i})X-X(%
\widehat{Q}_{N_{i}}^{i}+\widehat{Q}_{\pi _{i}}^{i})(XG_{0}+1)  \nonumber \\
&&-XG_{0}(\widehat{Q}_{N_{i}}^{i}+\widehat{Q}_{\pi _{i}}^{i})X\,\,\,+X(%
\widehat{Q}_{N_{f}}^{i}+\widehat{Q}_{\pi _{f}}^{i})G_{0}X\,\,\,  \nonumber \\
&=&(\widehat{Q}_{N_{f}}^{i}+\widehat{Q}_{\pi _{f}}^{i})X-X(\widehat{Q}%
_{N_{i}}^{i}+\widehat{Q}_{\pi _{i}}^{i})\,\,\,,\,  \label{kX}
\end{eqnarray}
where the superscript $i$ stands for ``intermediary.'' The result of $k_{\mu
}G_{0}g^{\mu }G_{0}$ was obtained from Eq.\ (\ref{eqg0}) of the Appendix. In
the first term, we have assumed here that what we want to prove holds true,
so again we only show self-consistency. Note that the structure of this
result is exactly the same as (\ref{Ucont3}), i.e., $X^{\mu }$ also
satisfies a continuity equation of the desired form. The final steps of the
calculation of $k_{\mu }U_{1}^{\mu }$ proceed now exactly analogous to (\ref
{kU0}) and we do not present the details here. The result is indeed 
\begin{equation}
k_{\mu }U_{1}^{\mu }=(\widehat{Q}_{N_{f}}+\widehat{Q}_{\pi _{f}})U_{1}-U_{1}(%
\widehat{Q}_{N_{i}}+\widehat{Q}_{\pi _{i}})\,\,\,\,,
\end{equation}
as stipulated.

We refrain from continuing this any further and simply mention that any
individual term of the nonpolar driving term $U$ of Eq.\ (\ref{eq25}) will
give rise to a current contribution satisfying a continuity equation of the
required structure. We conclude, therefore, that (\ref{Ucont3}) is true ---
i.e., is self-consistent with the general formalism developed here --- and
that the conjecture of the preceding subsection was justified.

%%%%%%%%%%%%%%%%%%%%%%%%%%%%%%%%%%%%%%%%%%%%%%%%%%%%%%%%%%%%%%%%%%%%%%%%

\subsection{Self-consistency of the Ward-Takahashi identities}

The final step to come full circle now in our presentation is to show that
the divergence of the dressed nucleon current $J^{\mu }$ of Eq.\ (\ref{eq28})
will give us back the Ward-Takahashi identity (\ref{WTN}). Using the same
commutator notation as in the previous step, we find for (\ref{eq28}):

\begin{eqnarray}
k_{\mu }J^{\mu } &=&S_{0}^{-1}\widehat{Q}_{N}-\widehat{Q}_{N}S_{0}^{-1}+%
\left( \widehat{Q}_{N}\left\langle F\right| -\left\langle F\right| \widehat{Q%
}^{i}\right) G_{0}\left| \Gamma \right\rangle +\left\langle \Gamma \right|
G_{0}\left( \widehat{Q}^{i}\left| F\right\rangle -\left| F\right\rangle 
\widehat{Q}_{N}\right)  \nonumber \\
&&+\left\langle \Gamma \right| \left( \widehat{Q}^{i}G_{0}-G_{0}\widehat{Q}%
^{i}\right) \left| \Gamma \right\rangle +\left\langle \Gamma \right|
G_{0}\left( \widehat{Q}^{i}U-U\widehat{Q}^{i}\right) G_{0}\left| \Gamma
\right\rangle  \nonumber \\
&=&\left( S_{0}^{-1}-\left\langle \Gamma \right| G_{0}\left| F\right\rangle
\right) \widehat{Q}_{N}-\widehat{Q}_{N}\left( S_{0}^{-1}-\left\langle
F\right| G_{0}\left| \Gamma \right\rangle \right)  \nonumber \\
&&-\left\langle F\right| G_{0}\widehat{Q}^{i}\left| \Gamma \right\rangle
+\left\langle \Gamma \right| G_{0}\widehat{Q}^{i}\left( \left|
F\right\rangle +UG_{0}\left| \Gamma \right\rangle \right)  \nonumber \\
&&+\left\langle \Gamma \right| \widehat{Q}^{i}G_{0}\left| F\right\rangle
-\left( \left\langle \Gamma \right| G_{0}U+\left\langle F\right| \right) 
\widehat{Q}^{i}G_{0}\left| \Gamma \right\rangle  \nonumber \\
&=&S^{-1}\widehat{Q}_{N}-\widehat{Q}_{N}S^{-1}\,\,\,  \label{kJpr}
\end{eqnarray}
which is exactly Eq.\ (\ref{qcN}). The charge operator $\widehat{Q}^{i}$
within the loop is the sum of the corresponding pion and nucleon charge
operators.

We thus have completed the proof of self-consistency of the present
formalism. \narrowtext

%%%%%%%%%%%%%%%%%%%%%%%%%%%%%%%%%%%%%%%%%%%%%%%%%%%%%%%%%%%%

\section{The continuity equations for the contact term}

Within the purely hadronic approach, i.e., without a detailed picture of the
underlying QCD reactions, all current mechanisms contributing to $F^{\mu }$
must come from interactions related to the three legs of the vertex since
there is no ``inside'' for a bare vertex. If this intuitive picture is to be
correct one should be able to read this off the expressions for the current $%
F^{\mu }$. To this end let us write (\ref{Gzero}) as

\begin{eqnarray}
\left| \widetilde{F}\right\rangle &=&\left| F_{p}\tau \right\rangle
Q_{N}-Q_{N}\left| F_{p}\tau \right\rangle -Q_{\pi }\left| F_{p}\tau
\right\rangle  \nonumber \\
&=&\left( \tau Q_{N}-Q_{N}\tau -Q_{\pi }\tau \right) \left|
F_{p}\right\rangle \,\,\,  \label{Gzero1}
\end{eqnarray}
which indeed is zero because of charge conservation at the bare vertex (the $%
Q$'s here have no hats), i.e., \widetext
\begin{equation}
\left( \tau Q_{N}-Q_{N}\tau -Q_{\pi }\tau \right) _{i}=\tau _{i}\frac{e}{2}%
(1+\tau _{3})-\frac{e}{2}(1+\tau _{3})\tau _{i}-ei\varepsilon _{i3j}\tau
_{j}=0\,\,\,;
\end{equation}
\narrowtext
\noindent in other words, we relate the validity of (\ref{Gzero}) to the
most basic conservation law available within the present context.
Subtracting then (\ref{Gzero1}) from both sides of Eq.\ (\ref{Gdef}), we are
led to define reaction changes for each leg of the bare vertex by

\begin{mathletters}
\label{rhob}
\begin{eqnarray}
\Delta _{k}R_{N_{i}} &=&\left( \left| F[q,p^{\prime },p+k]\right\rangle
-\left| F[q,p^{\prime },p]\right\rangle \right) \tau Q_{N}\,\,\,,
\label{rhob-a} \\
\Delta _{k}R_{N_{f}} &=&\left( \left| F[q,p^{\prime },p]\right\rangle
-\left| F[q,p^{\prime }-k,p]\right\rangle \right) Q_{N}\tau \,\,\,,
\label{rhob-b} \\
\Delta _{k}R_{\pi } &=&\left( \left| F[q,p^{\prime },p]\right\rangle -\left|
F[q-k,p^{\prime },p]\right\rangle \right) Q_{\pi }\tau \,\,\,,\,\,
\label{rhob-c}
\end{eqnarray}
\end{mathletters}
where the indices $N_{i}$ and $N_{f}$ denote the initial and final nucleons,
respectively, and $\pi $ the (final) pion. The notation chosen here for the
vertex functions exhibits all hadronic momenta, i.e.,

\begin{equation}
F=F[q_{\pi },p_{f},p_{i}]\,\,\,,
\end{equation}
with $p_{i}$, $p_{f}$ and $q_{\pi }$ the initial and final nucleon and pion
momenta, respectively, at the vertex (which are {\it not} necessarily the
corresponding momenta of the reaction). However, the choice of brackets $%
[...]$ rather than parentheses $(...)$ signifies that only the two
independent momenta are active. The third is silent in the sense that the
physically relevant vertex is given by a ray on which the dependent momentum
can take any value. For example, if we choose --- as we have done throughout
this paper --- the nucleon momenta as independent variables, then

\begin{equation}
F_{physical}\equiv F(p_{f},p_{i})=F[q_{\pi },p_{f},p_{i}]=F[q_{\pi }^{\prime
},p_{f},p_{i}]\,\,\,,
\end{equation}
irrespective of the value of $q_{\pi }$ specified in $F[q_{\pi
},p_{f},p_{i}] $; the physically relevant pion momentum for the vertex is $%
q=p_{i}-p_{f}.$ In other words, in writing down Eqs.\ (\ref{rhob}), we do 
{\it not} want to imply that $F[q,p^{\prime },p]$ is an unphysical vertex
here, which would violate momentum conservation,

\begin{equation}
p+k=p^{\prime }+q\,\,\,,
\end{equation}
in the pion production reaction.\footnote{%
This is rather different from the treatment of Ohta \cite{ohta}, who
requires unphysical values for vertex functions to assure gauge invariance.}
We merely want to leave the choice of independent variables open. As a
consequence, of course, once a choice has been made, one of the reaction
changes (\ref{rhob}) becomes identically zero. For example, if the nucleon
momenta are independent, then 
\begin{equation}
F[q,p^{\prime },p]-F[q-k,p^{\prime },p]=F(p^{\prime },p)-F(p^{\prime },p)=0
\end{equation}
and hence 
\begin{equation}
\Delta _{k}R_{\pi }=0\,\,\,.
\end{equation}
However, as we shall see, the contributions from hadron legs for which $%
\Delta _{k}R=0$ will enter through another mechanism as a matter of course.

With these preliminaries and without loss of generality, we can now write
the contact term as a sum of three currents, one for each leg of the vertex,
i.e.,

\widetext
\begin{equation}
\left| F_{p+k,p}^{\mu }\right\rangle =F^{\mu }(k;q,p^{\prime
},p)=j_{c,N_{i}}^{\mu }(p+k,p)\,+\,j_{c,N_{f}}^{\mu }(p^{\prime },p^{\prime
}-k)+\,j_{c,\pi }^{\mu }(q,q-k)\,\,\,,
\end{equation}
with each current satisfying a continuity equation,

$\,$

\begin{mathletters}
\label{bcont}
\begin{eqnarray}
k_{\mu }j_{c,N_{i}}^{\mu }+\Delta _{k}R_{N_{i}} &=&0\,\,\,,  \label{bcont-a}
\\
k_{\mu }j_{c,N_{f}}^{\mu }+\Delta _{k}R_{N_{f}} &=&0\,\,\,,  \label{bcont-b}
\\
k_{\mu }j_{c,\pi }^{\mu }+\Delta _{k}R_{\pi } &=&0\,\,\,.  \label{bcont-c}
\end{eqnarray}
\end{mathletters}
In other words, the validity of (\ref{Gzero}) is being assured in terms of
separate continuity equations for the three legs of the vertex. At this
stage, {\it any} choice of currents $j_{c}^{\mu }$ which satisfy these
continuity equations will lead to a gauge-invariant pion photoproduction
amplitude.

To see how this might work out in practice for pseudoscalar and pseudovector
couplings, let us consider the vertex of Eq.\ (\ref{eq12}) described in terms
of nucleonic momenta $p$ and $p^{\prime }$. Writing the coupling operator as

\begin{equation}
g_{0}G_{5}=G_{5ps}+G_{5pv}\frac{{q\!\!\!/}}{2m}\,\,\,,
\end{equation}
where $q$ is the appropriate pion momentum, we can treat both pseudoscalar ($%
G_{5ps}=g_{0}\gamma _{5}$, $G_{5pv}=0$) and pseudovector ($%
G_{5pv}=g_{0}\gamma _{5}$, $G_{5ps}=0$) couplings at the same time. In view
of our choice of independent momenta, the pion contact current and its
photonic reaction change vanish identically, i.e., $j_{c,\pi }^{\mu }=0$ and 
$\Delta _{k}R_{\pi }=0$. For the initial nucleon we have

\begin{eqnarray}
\Delta _{k}R_{N_{i}} &=&\left[ F(p^{\prime },p+k)-F(p^{\prime },p)\right]
\tau Q_{N}\,  \nonumber \\
&=&\left[ \left( G_{5ps}+G_{5pv}\frac{{p\!\!\!/}+{k\!\!\!/}-{p\!\!\!/}%
^{\prime }}{2m}\right) f(p^{\prime },p+k)-\left( G_{5ps}+G_{5pv}\frac{{%
p\!\!\!/}-{p\!\!\!/}^{\prime }}{2m}\right) f(p^{\prime },p)\right] \tau
Q_{N}\,  \nonumber \\
&=&\left( \widetilde{G}_{5}\left[ f(p^{\prime },p+k)-f(p^{\prime },p)\right]
+G_{5pv}\frac{{k\!\!\!/}}{2m}f(p^{\prime },p+k)\right) \tau Q_{N}\, 
\nonumber \\
&=&k_{\mu }\left[ \widetilde{G}_{5}\frac{n_{i}^{\mu }}{n_{i}\cdot k}\left[
f(p^{\prime },p+k)-f(p^{\prime },p)\right] +G_{5pv}\frac{\gamma ^{\mu }}{2m}%
f(p^{\prime },p+k)\right] \tau Q_{N}\,\,\,\,,  \label{rhoNi}
\end{eqnarray}
where 
\begin{equation}
\widetilde{G}_{5}=G_{5ps}+G_{5pv}\frac{{p\!\!\!/}-{p\!\!\!/}^{\prime }}{2m}%
\,\,\,,
\end{equation}
and, in the last step, we have introduced an arbitrary four-vector $%
n_{i}^{\mu }$ in order to be able to pull out an overall factor of $k_{\mu }$%
. This procedure is well-defined since $f(p^{\prime },p+k)-f(p^{\prime },p)$
compensates the singularity which would otherwise occur for vanishing photon
momentum. The most straightforward choice for $n_{i}^{\mu }$ is

\begin{equation}
n_{i}^{\mu }=(2p+k)^{\mu }
\end{equation}
since $p$ and $k$ are the only available four-vectors and $n_{i}^{\mu }$ it
is the well-known current operator of a scalar particle with momentum $p$.
The appropriateness of this choice is reinforced by noting that

\begin{equation}
\frac{n_{i}^{\mu }}{n_{i}\cdot k}=\frac{1}{2p\cdot k+k^{2}}(2p+k)^{\mu }=%
\frac{1}{(p+k)^{2}-p^{2}}(2p+k)^{\mu }\,\,\,,
\end{equation}
which is the dynamical picture of a scalar-particle electromagnetic vertex
and a subsequent scalar propagation with momentum $p+k$ and dynamical
``squared mass'' $p^{2}$ that indeed becomes $m^{2}$ when the external
nucleon is on-shell

The minimal choice for $j_{c,N_{i}}^{\mu }$ suggested by these
considerations, therefore, is simply 
\begin{equation}
j_{c,N_{i}}^{\mu }(p+k,p)=-\left( \widetilde{G}_{5}\frac{n_{i}^{\mu }}{%
n_{i}\cdot k}\left[ f(p^{\prime },p+k)-f(p^{\prime },p)\right] +G_{5pv}\frac{%
\gamma ^{\mu }}{2m}f(p^{\prime },p+k)\right) \tau Q_{N}\,\,\,\,.
\end{equation}
Similarly, for $j_{c,N_{f}}^{\mu }$ one finds 
\begin{equation}
j_{c,N_{f}}^{\mu }(p^{\prime },p^{\prime }-k)=-\left( \widetilde{G}_{5}\frac{%
n_{f}^{\mu }}{n_{f}\cdot k}\left[ f(p^{\prime },p)-f(p^{\prime }-k,p)\right]
+G_{5pv}\frac{\gamma ^{\mu }}{2m}f(p^{\prime }-k,p)\right) Q_{N}\tau
\,\,\,\,\,,
\end{equation}
with 
\begin{equation}
n_{f}^{\mu }=(2p^{\prime }-k)^{\mu }
\end{equation}
being the appropriate scalar operator.

When adding up the various contributions to obtain the contact current, one
has

\begin{mathletters}
\label{Fmudef}
\begin{eqnarray}
F^{\mu }(k;p^{\prime },p) &=&-\left[ \widetilde{G}_{5}\frac{n_{i}^{\mu }}{%
n_{i}\cdot k}\left[ f(p^{\prime },p+k)-f(p^{\prime },p)\right] +G_{5pv}\frac{%
\gamma ^{\mu }}{2m}f(p^{\prime },p+k)\right] \tau Q_{N}  \nonumber \\
&&-\left[ \widetilde{G}_{5}\frac{n_{f}^{\mu }}{n_{f}\cdot k}\left[
f(p^{\prime },p)-f(p^{\prime }-k,p)\right] -G_{5pv}\frac{\gamma ^{\mu }}{2m}%
f(p^{\prime }-k,p)\right] Q_{N}\tau  \label{Fmudef1} \\
&=&-G_{5pv}\frac{\gamma ^{\mu }}{2m}\left[ \tau Q_{N}\,f(p^{\prime
},p+k)-Q_{N}\tau \,f(p^{\prime }-k,p)\right]  \nonumber \\
&&-\widetilde{G}_{5}\left[ \,f(p^{\prime },p+k)\,\widehat{n}_{i}^{\mu }-%
\widehat{n}_{f}^{\mu }\,\,f(p^{\prime }-k,p)-\widehat{\pi}^{\mu
}\,f(p^{\prime },p)\right] \,\,\,.  \label{Fmudef2}
\end{eqnarray}
\end{mathletters}
\narrowtext
\noindent The first part of these expressions for $F^{\mu }$, Eq.\ (\ref
{Fmudef1}), is actually to be used in practical calculations --- since it is
explicitly free of any singularities at $k=0$. The second part, Eq.\ (\ref
{Fmudef2}), was written merely to exhibit the general structure of the
result. It shows that, apart from the spin-1/2 $\gamma ^{\mu }$ nucleon
pieces arising only in the pseudovector case, one has three scalar
contributions --- one for each leg, where the corresponding bare cutoff
function $f$ is multiplied by one of the following operators:

\begin{mathletters}
\label{iso}
\begin{eqnarray}
\widehat{n}_{i}^{\mu } &=&\frac{n_{i}^{\mu }}{n_{i}\cdot k}\,\tau Q_{N}\,=%
\frac{\tau Q_{N}}{(p+k)^{2}-p^{2}}(2p+k)^{\mu }\,\,,  \label{iNi} \\
\widehat{n}_{f}^{\mu } &=&\frac{n_{f}^{\mu }}{n_{f}\cdot k}\,Q_{N}\tau
=(2p^{\prime }-k)^{\mu }\frac{Q_{N}\tau }{(p^{\prime }-k)^{2}-p^{\prime 2}}%
\,\,\,,  \label{iNf} \\
\widehat{\pi}^{\mu } &=&\frac{n_{i}^{\mu }}{n_{i}\cdot k}\,\tau Q_{N}-\frac{%
n_{f}^{\mu }}{n_{f}\cdot k}\,Q_{N}\tau \,\,\,.  \label{ipi}
\end{eqnarray}
\end{mathletters}
Whereas the isospin description of the nucleons is static here, the last
operator, $\widehat{\pi}^{\mu }$, corresponds to a dynamic treatment in the
sense that the pion's isospin is obtained directly only upon taking the
divergence, i.e.,

\begin{equation}
k_{\mu }\widehat{\pi}^{\mu }=\tau Q_{N}-Q_{N}\tau =Q_{\pi }\tau \,\,\,.
\end{equation}
The reason for this is our present choice of taking both nucleon variables
as independent. If we had chosen a nucleon and a pion momentum as
independent, then the isospin descriptions of the pion would be analogous to
the nucleons' now and the isospin of the corresponding other nucleon would
change in analogy to Eq.\ (\ref{ipi}).

The result we have obtained here for the bare contact current $F^{\mu }$
certainly is not the most general form one can write down. One can add
arbitrary transverse pieces to the current (\ref{Fmudef}) without affecting
any of the gauge-invariance results. However, we would like to submit that
it is the simplest, non-trivial form that satisfies the requirement of
continuity equations that seems to govern every aspect of $\gamma \pi N$
physics. Moreover, since there does not yet exist a detailed derivation of
the bare hadron vertex from QCD, there is actually no dynamical basis for
going beyond the form proposed here.

The present expressions are reminiscent of Ohta's \cite{ohta} results for
extended nucleons obtained by minimal substitution and analytic continuation
(see also \cite{banerjee}). They are different, however, in two important
aspects. Since we formulate a vertex with arbitrary momenta not constrained
by momentum conservation in terms of rays rather than analytic continuation,
we do not require the cutoff function at unphysical values in the
expressions for the current, which considerably simplifies practical
applications. Furthermore, with our choice of independent momenta, an
explicit pion term is absent; rather, the pion's isospin is described
entirely in terms of nucleonic degrees of freedom. Overall, 
as shown by the form (\ref{Fmudef2}) of the 
contact current, with the operators defined as in (\ref{iso}), this 
has the advantage that one has a rather clear
interpretation of the underlying dynamical picture, with just one formfactor
per leg, with the appropriate momentum-conserving variable dependence.

%%%%%%%%%%%%%%%%%%%%%%%%%%%%%%%%%%%%%%%%%%%%%%%%%%%%%%%%%%%%%%%%%%%%%%%%

\section{Discussion}

We have presented here a complete and consistent description of the
interactions of pions, nucleons, and photons. It should be pointed out here
that the basic structure of the internal dynamical mechanisms for the pion 
photoproduction amplitude
obtained in Sec.\ III is the same as the one presented in the work of
van Antwerpen and Afnan \cite{afnan} (who used a different method of
derivation). As far as the final results are concerned,
the main difference is that these authors employ an expansion in terms of 
the irreducibility of the contributing mechanisms which seeks to avoid
nonlinearities in the final equations (see remarks before Eq.\ (3.30) of
\cite{afnan}) whereas we consider these nonlinearities an essential and
unavoidable consequence of the nature of the 
$\pi N$ and $\gamma \pi N$ systems. 
At the same time, however, the high degree of nonlinearity of our equations
presents the greatest practical obstacle to a numerical solution.
The nonlinearities occur at two stages. First, at
the purely hadronic level, in the way the full solution $X$ couples back
into the driving terms $U$, as described in Sec.\ II\@. Given the degree of
sophistication one wishes to achieve, there exist a number of obvious and
straightforward ways to approximate the hadronic driving term $U$ to render
the equations manageable from a practical point of view. Since this is not
our main concern, we will not enter a discussion here how this can be done
in detail. The second stage at which nonlinearities come to bear is at the
level of the electromagnetic interaction where the various pieces of the
current exhibit a high degree of nonlinear interdependence, as described in
Sec.\ III\@. Again, in practical calculations, one presumably needs to resort
to some approximations which --- at least partially --- linearize the
problem. The guiding principle for such approximations must be gauge
invariance. In other words, acceptable approximations of the currents should
at the very least maintain gauge invariance.

%%%%%%%%%%%%%%%%%%%%%%%%%%%%%%%%%%%%%%%%%%%%%%%%%%%%%%%%%%%%%%%%%%%%%%%%

\subsection{Approximating currents}

The considerations of the preceding two Secs.\ IV and V show that the gauge
invariance of all physical currents hinges on only one aspect of the
formalism. All current contributions resulting from the photon entering the
interior of the hadronic interaction region --- be it propagators, vertices
or other transition elements --- must satisfy continuity equations analogous
to those for the vertex current $\Gamma ^{\mu }$, the exchange current $%
U^{\mu }$, or the bare current $F^{\mu }$. In general, for every hadronic
reaction mechanism described by an operator $R$, with an associated
interaction current

\begin{equation}
R^{\mu }=-\{R\}^{\mu }\,\,\,,
\end{equation}
the quantity

\begin{equation}
\widetilde{R}=k_{\mu }R^{\mu }+R\,\widehat{Q}_{i}-\widehat{Q}_{f}R=0
\label{Ttilde}
\end{equation}
must vanish [cf. Fig.\ 8]. Here,

\begin{equation}
\widehat{Q}_{f}=\sum_{x_{f}}\widehat{Q}_{x_{f}}
\end{equation}
and

\begin{equation}
\widehat{Q}_{i}=\sum_{x_{i}}\widehat{Q}_{x_{i}}
\end{equation}
are the respective total charge operators for the final and initial channels
of the reaction, obtained by summing over the individual charge operators of
all outgoing or incoming legs. We recall that the operator $\widehat{Q}_{x}$
adds a photon momentum $k$ to particle $x$; with all external momenta given,
all momentum variables of $R$ are therefore unambiguously defined in Eq.\ (%
\ref{Ttilde}). Note that Eq.\ (\ref{Ttilde}) subsumes all continuity
equations considered so far, including the Ward-Takahashi identities [cf. (%
\ref{WT})].

Following the procedure of the last Section V concerning the bare current,
we may cast the condition that $\widetilde{R}$ vanish in the form 
\begin{equation}
\widetilde{R}=R_{P}\,Q_{i}-Q_{f}R_{P}=0\,\,\,,
\end{equation}
which is simply charge conservation. Note that the $Q$'s here have no hat,
and $P$ is the total momentum available for this reaction mechanism. Without
loss of generality, therefore, we may rewrite (\ref{Ttilde}) as

\begin{equation}
k_{\mu }R^{\mu }+(R\,\widehat{Q}_{i}-R_{P}\,Q_{i})+(Q_{f}R_{P}-\widehat{Q}%
_{f}R)=0\,\,\,.  \label{GaugeC}
\end{equation}
For the purpose of gauge invariance, it suffices now to approximate the full
current $R^{\mu }$ by 
\begin{equation}
R^{\mu }\rightarrow R_{approx}^{\mu }=\sum_{x_{f}}j_{c,x_{f}}^{\mu
}+\sum_{x_{i}}j_{c,x_{i}}^{\mu }\,\,\,,
\end{equation}
where each ``surface'' current $j_{c,x}^{\mu }$ satisfies an individual
continuity equation,

\begin{mathletters}
\label{kjr}
\begin{eqnarray}
k_{\mu }j_{c,x_{f}}^{\mu }+Q_{x_{f}}R_{P}-\widehat{Q}_{x_{f}}R &=&0\,\,\,,
\label{kjr1} \\
k_{\mu }j_{c,x_{i}}^{\mu }+R\,\widehat{Q}_{x_{i}}-R_{P}\,Q_{x_{i}}
&=&0\,\,\,\,,  \label{kjr2}
\end{eqnarray}
\end{mathletters}
which pertains only to a single charged leg of the reaction.

Let us illustrate this procedure for the exchange current $U^{\mu }$ of Sec.\
IV.B\@. One readily finds that the continuity equations for the individual
currents $j_{c}^{\mu }$ for each of the four legs of the driving term $U$
for $\pi N\rightarrow \pi N$ are given by

\begin{mathletters}
\label{kj}
\begin{eqnarray}
k_{\mu }j_{c,N_{f}}^{\mu }+Q_{N_{f}}\left[ U_{P}(p^{\prime
},p)-U_{P}(p^{\prime }-k,p)\right] &=&0\,\,\,,  \label{kj1} \\
k_{\mu }j_{c,\pi _{f}}^{\mu }+Q_{\pi _{f}}\left[ U_{P}(p^{\prime
},p)-U_{P}(p^{\prime },p)\right] &=&0\,\,\,,  \label{kj2} \\
k_{\mu }j_{c,N_{i}}^{\mu }+\left[ U_{P+k}(p^{\prime },p+k)-U_{P}(p^{\prime
},p)\right] Q_{N_{i}} &=&0\,\,\,,  \label{kj3} \\
k_{\mu }j_{c,\pi _{i}}^{\mu }+\left[ U_{P+k}(p^{\prime },p)-U_{P}(p^{\prime
},p)\right] Q_{\pi _{i}} &=&0\,\,\,.  \label{kj4}
\end{eqnarray}
\end{mathletters}
The independent momentum variables appearing here as arguments of $U$ are
those of the nucleons and the subscript $P$, or $P+k$, denotes the total
momentum available for the hadronic transition $U$; one has

\begin{equation}
P+k=p+q_{\pi _{i}}+k=p^{\prime }+q_{\pi _{f}}\,\,\,,
\end{equation}
where the $q_{\pi }$'s are the (suppressed) dependent pion momenta of the
exchange current $U_{P+k,P}^{\mu }(k;p^{\prime },p)$. Again, as with the
bare current in Sec.\ V, we find that choosing the nucleon momenta and the
total momentum as independent makes one of the pion-leg currents vanish,
namely (\ref{kj2}). Following the exact same procedure of Sec.\ V, we can
therefore approximate $U^{\mu }$ by the sum of four contact currents, one
for each incoming and outgoing leg,

\begin{equation}
U^{\mu }\rightarrow U_{approx}^{\mu }=j_{c,N_{f}}^{\mu }+j_{c,\pi _{f}}^{\mu
}+j_{c,N_{i}}^{\mu }+j_{c,\pi _{i}}^{\mu }\,\,\,\,,  \label{Umuj}
\end{equation}
with

\widetext
\begin{mathletters}
\label{jcex}
\begin{eqnarray}
j_{c,N_{f}}^{\mu } &=&-\frac{(2p^{\prime }-k)^{\mu }\,Q_{N_{f}}}{(2p^{\prime
}-k)\cdot k}\left[ U_{P}(p^{\prime },p)-U_{P}(p^{\prime }-k,p)\right] \,\,\,,
\label{jcex1} \\
j_{c,\pi _{f}}^{\mu } &=&0\,\,\,,  \label{jcex2} \\
j_{c,N_{i}}^{\mu } &=&-\left[ U_{P+k}(p^{\prime },p+k)-U_{P}(p^{\prime
},p)\right] \frac{Q_{N_{i}}\,(2p+k)^{\mu }}{(2p+k)\cdot k}\,\,\,,
\label{jcex3} \\
j_{c,\pi _{i}}^{\mu } &=&-\left[ U_{P+k}(p^{\prime },p)-U_{P}(p^{\prime
},p)\right] \frac{Q_{\pi _{i}}\,(2P-2p+k)^{\mu }}{(2P-2p+k)\cdot k}\,\,\,\,.
\label{jcex4}
\end{eqnarray}
\end{mathletters}
\narrowtext
\noindent This result assumes a pseudoscalar $\pi N$ coupling; for
pseudovector coupling, one might have additional terms involving $\gamma
^{\mu }$, similar to the findings of Sec.\ V. Note also that in the sum (\ref
{Umuj}), we can rearrange the terms such that the three contributions
containing $U_{P}(p^{\prime },p)$ provide a dynamic description of the
isospin of the final pion whose surface current $j_{c,\pi _{f}}^{\mu }$ (\ref
{jcex2}) vanishes here due to our choice of independent variables. The
considerations for the bare current of the preceding Sec.\ V regarding a
dynamic isospin description for the pion thus carry over to the outgoing
pion leg of the exchange current in complete analogy.

The single-particle surface currents (\ref{jcex}) are sufficient to provide
the same gauge condition as the exact exchange current $U^{\mu }$, 
\begin{equation}
k_{\mu }U^{\mu }=k_{\mu }U_{approx}^{\mu }\,\,\,.
\end{equation}
and thus allow one to preserve gauge invariance without having to take into
account {\it any} of the complex mechanisms contributing to $U^{\mu }$. One
could, of course, go to a more sophisticated (and more complicated) level of
approximation if, instead of for $U^{\mu }$ directly, one employed similar
surface current approximations for the current ingredients that contribute
to $U^{\mu }$ [cf. Eqs.\ (\ref{eq33-2})-(\ref{eq35})]. In other words,
depending on the hierarchical level of the reaction mechanism at which one
employs the approximation scheme presented here, one has complete control
over the degree of sophistication without ever sacrificing gauge invariance.

%%%%%%%%%%%%%%%%%%%%%%%%%%%%%%%%%%%%%%%%%%%%%%%%%%%%%%%%%%%%%%%%%%%%%%%%

\subsection{Summary}

Despite the fact that the nonlinear formalism presented here is extremely
complex in its full implementation, it is quite simple as far as its general
structure is concerned. As we hope to have made clear, it lends itself
immediately to approximations which can be as cursorily or as detailed as
desired. Following the general procedure outlined above, gauge invariance is
never at issue, since the exact gauge condition can always be turned into a
set of single-particle ``surface'' continuity equations for all charged
particles entering or leaving the interaction region. It is obvious that
this will remain true even if applied to other mechanisms, for example, eta
photoproduction, since they can be treated in complete analogy to the
present formalism.

Furthermore, it is equally obvious that the present photoproduction
formalism also carries over to larger hadronic systems since formally every
many-body (or even infinite-body) problem can be turned into an effective
scattering problem similar in structure to Fig.\ 1(c) with all complicated
subsystem reaction mechanisms subsumed into a driving term not unlike the
structure found for $U$ here (see Fig.\ 2) \cite{hh}. The gauge conditions
for this larger hadronic system, therefore, look similar to what we have
found here for $\pi N$, and the same type of approximations will allow one
to preserve gauge invariance.

In summary, the formalism developed here provides not only a detailed
dynamical picture of all contributing reaction mechanisms but at the same
time suggests a consistent approximation scheme that allows one to ensure
the gauge invariance of the final result. At whatever level of the reaction
dynamics one chooses to employ this scheme, the required pieces involve only
purely hadronic contributions, with the photon's effect on the system being
described simply by the change brought about in the hadronic reaction by
feeding an extra photon momentum through the interaction region but
otherwise leaving the hadronic mechanisms undisturbed. This provides an
intuitively appealing and practically easily manageable way of maintaining
gauge invariance for all interactions of photons with hadrons.

%===========================================================================

\acknowledgments

The author would like to thank Professor J. Speth and the staff of the
Institut f\"{u}r Kernphysik at the Forschungszentrum J\"{u}lich for their
hospitality during his recent stay when part of the work reported here was
done. This work was supported by Grant No. DE-FG02-95ER40907 of the 
U.S. Department of Energy.

%%%%%%%%%%%%%%%%%%%%%%%%%%%%%%%%%%%%%%%%%%%%%%%%%%%%%%%%%%%%%%%%%%%%%%%%

\appendix

\section*{Gauge derivative and currents defined}

The connected part of the
$n$-point Green's function of a hadronic transition described by an
amplitude $T_{P}$, with $m$ incoming and $n-m$ outgoing hadrons is given
schematically as

\begin{equation}
G_{T}=[t_{f_{1}}t_{f_{2}}...t_{f_{n-m}}]_{P}\,T_{P}%
\,[t_{i_{1}}t_{i_{2}}...t_{i_{m}}]_{P}\,\,\,,
\end{equation}
where $[t_{i_{1}}t_{i_{2}}...t_{i_{m}}]_{P}$ and $%
[t_{f_{1}}t_{f_{2}}...t_{f_{n-m}}]_{P}$ are the products of propagators of
all initial and final hadron legs, respectively, of the process; the index $%
P $ signifies the total momentum. Gauging the momenta appearing in $G_{T}$
according to the minimal-substitution rule,

\begin{equation}
p^{\mu }\rightarrow p^{\mu }-Q\,A^{\mu }\,\,,  \label{a1}
\end{equation}
the resulting Green's function, denoted symbolically by $G_{T,A}$, to first
order in the electromagnetic field $A^{\mu }$, becomes\footnote{%
This is merely a sketch of the procedure and of course not to be taken as a
rigorous derivation.}

\widetext
\begin{equation}
G_{T,A}\rightarrow
G_{T}+[t_{f_{1}}t_{f_{2}}...t_{f_{n-m}}]_{P+k}\,M_{T_{P+k,P}}^{\mu
}\,[t_{i_{1}}t_{i_{2}}...t_{i_{m}}]_{P}\,A_{\mu }\,\,\,,
\end{equation}
where $k$ is the momentum of the photon, and $M_{T_{P+k,P}}^{\mu }$ the
electromagnetic current associated with the hadronic transition $T_{P}$.
This result amounts to defining the current as 
\begin{equation}
M_{T_{P+k,P}}^{\mu }=[t_{f_{1}}t_{f_{2}}...t_{f_{n-m}}]_{P+k}^{-1}\,\left[ 
\frac{\delta }{\delta A_{\mu }}G_{T,A}\right] _{A_{\mu
}=0}[t_{i_{1}}t_{i_{2}}...t_{i_{m}}]_{P}^{-1}\,\,\,.  \label{Mfunc}
\end{equation}
In other words, currents are described by a
Lehmann--Symanzik--Zimmermann-type reduction procedure \cite{weinberg,LSZ}.

We would like to introduce an operation which achieves the same result yet
is very simple to use. Replacing the functional derivative in (\ref{Mfunc})
by 
\begin{equation}
\left[ \frac{\delta }{\delta A_{\mu }}\,\,G_{T,A}\right] _{A_{\mu
}=0}\rightarrow -\{G_{T}\}^{\mu }\,\,\,,
\end{equation}
i.e.,

\begin{equation}
M_{T_{P+k,P}}^{\mu }=-[t_{f_{1}}t_{f_{2}}...t_{f_{n-m}}]_{P+k}^{-1}\left\{
[t_{f_{1}}t_{f_{2}}...t_{f_{n-m}}]_{P}\,T_{P}%
\,[t_{i_{1}}t_{i_{2}}...t_{i_{m}}]_{P}\right\} ^{\mu
}[t_{i_{1}}t_{i_{2}}...t_{i_{m}}]_{P}^{-1}\,\,\,,  \label{Mmu}
\end{equation}
\narrowtext
\noindent we introduce an operation which we call a ``gauge derivative,''
denoted by the symbol \{...\}$^{\mu }$. It acts on the total-momentum
operator of the particular subsystem to which the photon attaches itself and
is defined by the following rules.

{\it Rule 0: } \{...\}$^{\mu }$ is a linear operation, i.e., with $\alpha $, 
$\beta $ being numbers, 
\begin{equation}
\{\alpha A+\beta B\}^{\mu }=\alpha \{A\}^{\mu }+\beta \{B\}^{\mu }\,\,\,,
\end{equation}
and its action on a constant produces a zero, 
\begin{equation}
\{{\rm const}\}^{\mu }=0\,\,\,  \label{r1}
\end{equation}
(because such an entity cannot absorb the photon's momentum).

{\it Rule 1: }The action on a momentum component $p^{\nu}$ 
produces the charge
operator $Q$ for the 
particle or system carrying momentum $p$ multiplied by the metric tensor, 
\begin{equation}
\{p^{\nu }\}^{\mu }=Qg^{\mu \nu }\,\,\,.  \label{r2}
\end{equation}
The origin of this rule is the functional derivative of the minimal
substitution (\ref{a1}), i.e.,

\begin{equation}
\{p^{\nu }\}^{\mu }\equiv -\frac{\delta }{\delta A_{\mu }}(p^{\nu }-QA^{\nu
})=Qg^{\mu \nu }\,\,\,,
\end{equation}
which is the reason for calling $\{...\}^{\mu }$ a gauge derivative. (Note
that this is the only place where the particular nature of the
electromagnetic field enters the rules. Other types of gauge fields would
produce a different result here.)

{\it Rule 2:} Any momentum-conserving delta function valid for all parts
within the gauge-derivative braces $\{...\}^{\mu }$ must be taken outside
the braces and replaced by one where the total initial momentum $p$ is
shifted by the photon momentum $k$, i.e., 
\begin{equation}
\{\delta (p^{\prime}-p)B(p^{\prime})A(p)\}^{\mu }=\delta (p^{\prime}-p-k)\{B(p^{\prime})A(p)\}^{\mu }\,\,\,.  \label{r3}
\end{equation}
In many instances, there will only be an implicit delta function because all
terms are already written taking into account momentum conservation. For
such cases, the result must be taken to have an overall implicit delta
function with shifted momenta. (Formally, however, one is on safer grounds
if one writes out all delta functions explicitly and then removes them after
having taken the gauge derivatives.)

{\it Rule 3: } 
\begin{equation}
\{B(p)A(p)\}^{\mu }=B(p+k)\{A(p)\}^{\mu }+\{B(p)\}^{\mu }A(p)\,\,\,.%  \label{r4}
\end{equation}
The physical background for this rule is that if the photon attaches itself
to a process described by two successive hadronic subprocesses $A$ and $B$
depending on the same (conserved) total four-momentum $p$, one can break up
the description by attaching it first to subprocess $A$ and then to $B$. In
the first case, however, the total four-momentum available for process $B$
has changed by the momentum $k$ of the photon. Note that this is an example
where there is an implicit delta function relating the momentum arguments of 
$B$ and $A$.

{\it Rule 4: }If there is a branch in a hadronic reaction where schematically

\[
A(p)\rightarrow B(p^{\prime })+C(q)\,\,\,, 
\]
with $p=p^{\prime }+q$ and $\Gamma (q,p^{\prime },p)$ describing the
transition, this rule states that \widetext
\begin{eqnarray}
\{B(p^{\prime })C(q)\Gamma (q,p^{\prime },p)\delta _{0}A(p)\}^{\mu }
&=&\delta _{k}\{B(p^{\prime })C(q)\Gamma (q,p^{\prime },p)A(p)\}^{\mu } 
\nonumber \\
&=&\delta _{k}B(p^{\prime })C(q)\Gamma (q,p^{\prime },p+k)\{A(p)\}^{\mu } 
\nonumber \\
&&+\delta _{k}B(p^{\prime })C(q)\{\Gamma (q,p^{\prime },p)\}^{\mu }A(p) 
\nonumber \\
&&+\delta _{k}B(p^{\prime })\{C(q-k)\}^{\mu }\Gamma (q-k,p^{\prime },p)A(p) 
\nonumber \\
&&+\delta _{k}\{B(p^{\prime }-k)\}^{\mu }C(q)\Gamma (q,p^{\prime }-k,p)A(p)%
\,\,\,, \label{r5}
\end{eqnarray}
\narrowtext
\noindent where $\delta _{0}=\delta (p-p^{\prime }-q)$ and $\delta
_{k}=\delta (p+k-p^{\prime }-q)$ abbreviate the delta functions. The
branching occurs in the last two terms where in the first the $C$-branch is
gauged and in the second the $B$-branch. This rule follows from momentum
conservation at the vertex $\Gamma (q,p^{\prime },p)$ and from the fact that
the external momenta are fixed already; it takes care also of loop processes 
$A\rightarrow B+C\rightarrow D$. Note that $B(p^{\prime })$ and $C(q)$ are
independent reaction mechanisms after the branching, tied together only by
momentum conservation; in other words, they correspond to a convolution $%
B\circ C$ similar to $G_{0}=S\circ \Delta $ of Eq.\ (\ref{eq14}).

Having established the rules, the current $J^{\mu }$ for a nucleon is given
via gauging the nucleon propagator $S$ (which is the appropriate two-point
Green's function for this case) and amputating the propagators of external
legs according to (\ref{Mmu}), i.e., 
\begin{equation}
J^{\mu }(p+k,p)=-S^{-1}(p+k)\{S(p)\}^{\mu }S^{-1}(p)\,\,\,.  \label{JN}
\end{equation}
Because of $\{SS^{-1}\}^{\mu }=\{1\}^{\mu }=0$ and Rule 3, we have 
\begin{equation}
J^{\mu }(p+k,p)=\{S^{-1}(p)\}^{\mu }\,\,\,  \label{Ncurr}
\end{equation}
as the general definition of the nucleon current operator. Similarly for the
pion,

\begin{equation}
J_{\pi }^{\mu }(q+k,q)=\{\Delta ^{-1}(q)\}^{\mu }\,\,\,.  \label{picurr}
\end{equation}
For the example of bare propagators, one then reproduces the expected
results for both the nucleon and the pion, i.e.,

\begin{eqnarray}
J_{0}^{\mu }(p+k,p) &=&\{S_{0}^{-1}(p)\}^{\mu }  \nonumber \\
&=&\{{p\!\!\!/}-m\}^{\mu }  \nonumber \\
&=&\{{p\!\!\!/}\}^{\mu }  \nonumber \\
&=&(p+k)^{\nu }\{\gamma _{\nu }\}^{\mu }+\{p^{\nu }\}^{\mu }\,\gamma _{\nu }
\nonumber \\
&=&Q_{N}\,\gamma ^{\mu }\,\,\,,  \label{a7}
\end{eqnarray}
and 
\begin{eqnarray}
J_{\pi 0}^{\mu }(q+k,q) &=&\{\Delta _{0}^{-1}(q)\}^{\mu }  \nonumber \\
&=&\{q^{2}-m_{\pi }^{2}\}^{\mu }  \nonumber \\
&=&\{q^{2}\}^{\mu }  \nonumber \\
&=&(q+k)^{\nu }\{q_{\nu }\}^{\mu }+\{q^{\nu }\}^{\mu }q_{\nu }  \nonumber \\
&=&Q_{\pi }(2q+k)^{\mu }\,\,\,;  \label{a8}
\end{eqnarray}
$Q_{N}$ and $Q_{\pi }$ are the respective charge operators.

For the pion-production current $M^{\mu }$ for a nucleon with momentum $p$
going into a nucleon and a pion with total momentum $p+k$ upon absorbing a
photon with momentum $k$, the definition (\ref{Mmu}) then yields

\begin{equation}
M^{\mu }(p+k,p)=-G_{0}^{-1}(p+k)\,\{G_{0}(p)\Gamma (p)S(p)\}^{\mu
}\,S^{-1}(p)\,\,,  \label{MCurr}
\end{equation}
where the quantity in the gauge-derivative braces is the three-point Green's
function for $N\rightarrow N+\pi $, with $\Gamma $ being the dressed vertex
of Eq.\ (\ref{eq22}). We have exhibited here only the total momentum of the
system; the details are to be found in Sec.\ III.

This completes the definitions of the gauge derivative and the currents. Let
us add a note of caution here. The gauge derivative is based on the
assumption that the quantities it acts on are physically meaningful in the
sense that they can be broken down into their reaction-dynamical content.
Its application, therefore, does not seem to be warranted when this is no
longer possible. An example of this is the application to the bare vertex $F$
of Eq.\ (\ref{eq12}),

\begin{equation}
F^{\mu }(k;p^{\prime },p)=-\{F(p^{\prime },p)\}^{\mu }\,.  \label{fbaremu}
\end{equation}
Since $F(p^{\prime },p)$ is an (at this stage largely ambiguous)
parametrization of the unsolved underlying QCD dynamics, it is in our
opinion not very meaningful to apply the procedure to the functional form of 
$F(p^{\prime },p)$. This is quite different from the corresponding quantity
of the dressed vertex [cf.\ Eq.\ (\ref{eq29}) and Fig.\ 5],

\begin{equation}
\Gamma ^{\mu }(k;p^{\prime },p)=-\{\Gamma (p^{\prime },p)\}^{\mu }\,,
\label{Gmu}
\end{equation}
which is based on the detailed dynamical picture developed here.

\widetext
Finally, since it is required in Secs.\ III and IV, let us look at letting $%
\{...\}^{\mu }$ act on the $\pi N$ propagator $G_{0}=\Delta \circ S$, 
\begin{eqnarray}
-\{G_{0}(p+q)\}^{\mu } &=&-\{\Delta (q)\}^{\mu }\circ S(p)-\Delta (q)\circ
\{S(p)\}^{\mu }  \nonumber \\
&=&\left[ \Delta (q+k)\{\Delta ^{-1}(q)\}^{\mu }\Delta (q)\right] \circ S(p)
+\left[ S(p+k)\{S^{-1}(p)\}^{\mu }S(p)\right] \circ \Delta (q)  \nonumber \\
&=&\left[ \Delta \,(q+k)J_{\pi }^{\mu }(q+k,q)\Delta (q)\right] \circ S(p)
+\left[ S(p+k)\,J^{\mu }(p+k,p)\,S(p)\right] \circ \Delta (q)  \nonumber \\
&=&G_{0}(p+q+k)g^{\mu }(k;p,q)G_{0}(p+q)\,\,\,,  \label{eqg0}
\end{eqnarray}
where $p+q$ and $p+q+k$ are the total initial and final hadronic
four-momenta; $G_{0}g^{\mu }G_{0}$ in the last step is only a convenient
short-hand notation defined by the preceding expression. If one removes the
left-hand side $G_{0}$ from $G_{0}g^{\mu }G_{0}$

\begin{equation}
g^{\mu }(k;p,q)G_{0}(p+q) =\left[ J_{\pi }^{\mu }(q+k,q)\,\Delta (q)\right]
\circ {\bf 1}_{N}+\left[ J^{\mu }(p+k,p)\,S(p)\right] \circ {\bf 1}_{\pi
}\,\,\,,  \label{gmug0}
\end{equation}
with ${\bf 1}_{N}$ and ${\bf 1}_{\pi }$ denoting momentum conservation for
the respective particles, and hence 
\begin{equation}
g^{\mu }(k;p,q) =J_{\pi }^{\mu }(q+k,q)\,\circ S^{-1}(p) +J^{\mu
}(p+k,p)\,\circ \Delta ^{-1}(q)\,\,\,.
\end{equation}
\narrowtext

%%%%%%%%%%%%%%%%%%%%%%%%%%%%%%%%%%%%%%%%%%

%%%%%%%%%%%%%%%%%%%%%%%%%%%%%%%%%%%%%%%%%%%

\end{document}